\documentclass[aps,pre,twocolumn,longbibliography]{revtex4-1}

\usepackage{amsmath,amssymb,amsfonts,hyperref}
\usepackage{graphicx,epsf}
\usepackage{xspace}
\usepackage{textcomp}
\usepackage{color}
\usepackage{verbatim}
\usepackage{multirow}
\usepackage{diagbox}
\usepackage{ulem}

\usepackage{abbrevs}
\newif\ifhyper
\hypertrue
\ifhyper
\hypersetup{
  citecolor = {green},
  urlcolor = {blue} 
} 

\newcommand\modif[1]{{#1}}

\newcommand\avg[1]{\left\langle{#1}\right\rangle}
\newcommand{\p}{\partial} 
\newcommand{\vx}{\vec{x}}

\newcommand{\vv}{\vec{v}}

\newcommand{\vnabla}{\vec{\nabla}}

\newcommand{\vf}{\vec{f}\,}

\newcommand{\vk}{{\vec{k}}}

\newcommand{\ie}{{\it i.e.}}

\newcommand{\sref}[1]{Sec.~\ref{#1}}
\newcommand{\fref}[1]{Fig.~\ref{#1}}
\newcommand{\tref}[1]{Tab.~\ref{#1}}
\newcommand{\aref}[1]{Appendix~\ref{#1}}

\newcommand{\Eq}[1]{Eq.~(\ref{#1})}
\newcommand{\eq}[1]{(\ref{#1})}

\usepackage{abbrevs}
\usepackage{etoolbox}
\newabbrev\NPRG{Non-perturbative Renormalisation Group (NPRG)}[NPRG]
\newabbrev\RG{Renormalization Group (RG)}[RG]
\newabbrev\NS{Navier-Stokes (NS)}[NS]
\newabbrev\threeD{three-dimensional (3D)}[3D]
\newabbrev\DNS{Direct Numerical Simulations (DNS)}[DNS]
\newabbrev\Relam{Taylor microscale Reynolds number (Re$_\lambda$)}[Re$_\lambda$]

\robustify{\NPRG}
\robustify{\RG}
\robustify{\NS}
\robustify{\DNS}
\robustify{\threeD}
\robustify{\Relam}

\makeatletter
\renewcommand\maybe@space@{%
  \maybe@ictrue 
  \expandafter   \@tfor
    \expandafter \reserved@a
    \expandafter :%
    \expandafter =%
                 \nospacelist
                 \do \t@st@ic
  \ifmaybe@ic 
    \space
  \fi
}
\makeatother

\begin{document}

\title{Analysis of the dissipative range of the energy spectrum in grid turbulence and in direct numerical simulations}

\author{Anastasiia Gorbunova$^{1,2}$, Guillaume Balarac$^{2,4}$, Micka\"el~Bourgoin$^3$, L\'eonie Canet$^{1,4}$, Nicolas Mordant$^{2}$, 
  Vincent Rossetto$^1$}
\affiliation{$^1$ Univ. Grenoble Alpes, CNRS, LPMMC, 38000 Grenoble, France\\
$^2$ Univ. Grenoble Alpes, CNRS, LEGI, 38000 Grenoble, France\\
$^3$ ENS Lyon and CNRS, Physics department, 69000 Lyon, France\\
$^4$ Institut Universitaire de France (IUF)}

\begin{abstract}
We present a statistical analysis of the behavior of the kinetic energy
spectrum in the dissipative range of fully developed three-dimensional 
turbulence, \modif{
	with the aim of testing a recent prediction obtained from the
non-perturbative renormalization group. Analyzing spectra
recorded in experiments of grid turbulence, generated in the Modane wind tunnel,
and spectra obtained from high-resolution direct numerical simulations of the
forced Navier-Stokes equation, we observe that the spectra decay as a stretched
exponential in the dissipative range. The theory predicts a stretching exponent~$\alpha=2/3$,
  and the data analyses of the numerical and experimental spectra are in close agreement with this value. }
This result also corroborates previous DNS
studies which found that the spectrum in the near-dissipative range is best
modeled by a stretched exponential with $\alpha<1$. 
\end{abstract}

\maketitle

\section{Introduction}

The very chaotic nature of the motion of a fluid driven to a turbulent state calls for a statistical description.
  A striking feature of \threeD turbulence is the emergence of very robust universal statistical properties,
  such as the well-known $k^{-5/3}$
 decay of the energy spectrum over a wide range of length scales called the inertial range. This inertial range exists
 at sufficiently high Reynolds numbers, 
 when the typical scale at which energy is injected (called the integral scale $L$) and the microscopic scale
   (called the Kolmogorov scale $\eta$) at which it is dissipated by molecular friction, are well separated.

 The first understanding of these properties was provided by the pioneering statistical theory of turbulence proposed
 by Kolmogorov, and referred to as K41 \cite{Kolmogorov41a,Kolmogorov91a,Kolmogorov41c,Kolmogorov91b}. 
  K41 theory relies on the fundamental assumptions that 
  the small-scale turbulence is statistically independent
 of the large scales, that it is locally homogeneous, isotropic and steady, and that the average
   energy dissipation rate per unit mass $\epsilon$ is finite and 
   independent of the kinematic viscosity $\nu$ in the limit $\nu \to 0$.
  This implies
  that the statistical properties of small-scale turbulence should be determined uniquely by  $\epsilon$
 and  $\nu$. In particular,  the energy spectrum should take  the universal  from
\begin{equation}
 E(k) = \epsilon^{2/3} k^{-5/3} f(k\eta)
\end{equation}
where $\eta=\nu^{3/4}\epsilon^{-1/4}$ is the Kolmogorov length scale.
 The function $f$ is universal. In the inertial range, viscous effects are expected to be negligible, so that the spectrum
  should not depend on $\eta$, which implies that $f(x)$  should tend to a constant $C_K$ for $x\to 0$, 
 while $f$ should  fastly decay at large wave-numbers 
 $x\gtrsim1$, which corresponds to the dissipative regime
\cite{Kraichnan59}.
  
Although $f$ is expected to be universal, its analytical expression is not known. 
  Several early empirical expressions of the form
 $f(x) \sim x^{-\beta} \exp(-\mu x^\alpha)$,  with different values for $\alpha$ (1/2 \cite{Tatarskii67}, 
 3/2 \cite{Uberoi69}, 4/3 \cite{Pao65} or 
  2 \cite{Townsend51b,Novikov61,Gurvich67}) were proposed, on the basis of approximate fits of experimental data 
 or (approximate) analytical considerations \cite{Monin73}.
 Besides, different theoretical arguments focusing on the limit of large $k$
 (Direct Interaction Approximation \cite{Kraichnan59}, asymptotic expansions  \cite{Foias90,Sirovich94,Lohse95})
  advocated that in this limit, {\it i.e.} sufficiently deep in the far-dissipative range,
  the spectrum should decay as a pure exponential $\alpha=1$. \modif{Alternatively, the multi-fractal formalism 
  predicts a universal form of the spectrum involving the Reynolds number \cite{Frisch91}}.
 
 Subsequently, the behavior of the spectrum in the dissipative range has been extensively studied in
  experiments \cite{Sreenivasan85,Smith91,She93,Saddoughi94} and \DNS  \cite{Sanada92,Chen93,Martinez97,Ishihara05,Schumacher07,Ishihara09,Verma18}. 
  Accurate experimental measurements of the spectrum at small scales are difficult to obtain, 
  and several fits of the spectrum in the dissipative range have been proposed,
  for instance pure exponential functions with $\alpha=1$ but on two successive separate ranges in \cite{Sreenivasan85},
   or a single pure exponential but with  different coefficients $\mu$ in \cite{Saddoughi94} and \cite{Manley92}.
   Conversely, the analysis of Ref. \cite{Smith91} concluded that an exponential with $\alpha=2$ was the best fit
    for the experimental data, while still another empirical form is proposed in \cite{Pope00}.
      
                  The \DNS can in principle provide more accurate data for the spectrum in the dissipative range,
         but  their  high computational cost limits the studies to either low or moderate Reynolds 
         numbers in order to reach high spectral resolution \cite{Chen93,Martinez97,Schumacher07,Verma18} or
          to lower spectral resolutions to reach higher Reynolds numbers
          \cite{Sanada92,Ishihara05,Ishihara09}.  In practice, most of these works  assumed that $\alpha=1$, and aimed
          at determining the exponent $\beta$ for the power
       laws (or combination  of power laws) in front and the value of $\mu$.
        Although a pure exponential fit was shown to be a reliable model for the spectrum at very low Reynolds numbers \cite{Verma18}, it turned out not to be suitable to model the whole dissipative range already at moderate Reynolds numbers  \cite{Martinez97}, and the result of the fit was found to depend significantly on the choice of the fitting range  \cite{Ishihara05}.

  An extended analysis of the existing results for the dissipative range is provided in a recent account \cite{Khurshid18}. 
  This work (and previous studies) point to the conclusion that there exist two distinct regimes:
  the near-dissipative range (NDR) for $0.2 \lesssim k\eta \lesssim 4$ and the far-dissipative range (FDR)
   for $k\eta \gtrsim 4$. In the NDR, the
    logarithmic derivative of the spectrum is not linear, such that a pure exponential is not a consistent description,
     and its curvature indicates that $\alpha<1$.     In the FDR, 
     the decay of the spectrum is well described by an exponential ($\alpha=1$). 
   \modif{This finding explains the failure of previous attempts to describe the whole dissipative range
       as a pure exponential with various power-laws. The authors of Ref.~\cite{Khurshid18}
       proposed a phenomenological modeling as a superposition of two exponentials, one with $\alpha<1$
       dominating in the NDR  and one with $\alpha>1$ dominating in the FDR.}      
       Note that no consensus seems to be reached 
  concerning the power-laws multiplying the exponentials in either regimes.

   Independently, a theoretical prediction for the behavior of the spectrum
  in the dissipative range has been recently obtained from a \NPRG approach. 
  This theoretical approach is a ``first-principles'' one, in the sense that it is based on the  Navier-Stokes equation, without  involving any phenomenological inputs nor uncontrolled approximations.
 It has recently led to a progress in the understanding of homogeneous isotropic and stationary  turbulence,
 by providing the  time dependence of multi-point correlation functions in the turbulent state \cite{Tarpin18,Tarpin19}. 
  Concerning the energy spectrum, the NPRG  yields
 a stretched exponential behavior with $\alpha=2/3$ in the NDR,  and a regular one (pure exponential) in the FDR. 
 \modif{This prediction provides  a theoretical justification for the two-exponential phenomenological model
  proposed in~\cite{Khurshid18}. }
  This value for $\alpha$ in the NDR was then confirmed in \DNS \cite{Canet17}, and also
  in experiments of von K\'arm\'an turbulent swirling flow \cite{Debue18}.  However, 
  obtaining  a quantitative estimate of the  exponent of the stretched exponential
   is a difficult task, as it requires a sufficiently extended dissipative range and a high resolution. 
   
   In this work, we perform \DNS with a different compromise compared to previous studies
    favoring  higher  Reynolds numbers while keeping a sufficient spectral resolution in the NDR in order to 
   reliably probe this regime. Moreover, we present a statistical analysis of the experimental data recorded 
  in the Modane wind tunnel, featuring grid turbulence. Grid turbulence appears 
  as a particularly suitable set-up to investigate the dissipative range, since high Reynolds numbers
  can be attained, and the Kolmogorov scale is typically larger
 in the air than in liquids, such
  that a higher resolution can be expected. 
 Besides, the unique dimensions of the S1MA wind tunnel of ONERA in Modane, where the experiments discussed here were performed, allow one to investigate relatively high Reynolds number regimes with yet experimentally well resolved dissipative scales \cite{Bourgoin17}.
 In the following, we focus on the NDR, and more specifically on the exponential part, irrespective of the power-laws.
  We find from our analysis that the energy spectrum follows the predicted stretched exponential behavior,
   with an estimated  exponent $\alpha = 0.68 \pm 0.19$, in close agreement with the NPRG result.
 
  The remainder of the paper is organized as follows: in \sref{sec:theory}, we give some details on the \NPRG
   prediction. We present in \sref{sec:num} the results from the  \DNS,
   and in \sref{sec:exp} the statistical analysis of the experimental data. Additional details are reported in the Appendices.

\section{Theoretical predictions from the non-perturbative Renormalization Group}
\label{sec:theory}

We present in this section the principle of the \NPRG and the main ideas 
underlying the derivation of
\Eq{spectre} for the spectrum in the dissipative range, 
which we aim at testing. 
The purpose of this section is not to give technical details on its derivation, 
 which can be found in Refs. \cite{Canet17,Tarpin18}, but rather
 to emphasize the underlying assumptions: it is based on the stochastic Navier-Stokes
  equation, and on an expansion at large wave-numbers, which leading term can be calculated
   exactly without any further approximations. One may resume directly at \Eq{eq:TF C} for the result relevant for this work.

 The idea of applying \RG to turbulence 
\cite{Dominicis79,Fournier83,Smith98,Adzhemyan99,Zhou10} originates in the
observation that the statistical properties of a turbulent flow are
universal and described by power-laws in the inertial range. These power-laws are the hallmarks of scale invariance. 
The source of scale invariance at a critical point is  
the emergence of fluctuations at all scales. 
The \RG, as originally conceived by Wilson \cite{Wilson74},
is a method to efficiently average over these (non-Gaussian) fluctuations
to obtain the critical properties of the system. 
It should thus provide a valuable tool  to study fully developed
turbulence, which intrinsically involves length scales over many orders
of magnitude.

The \NPRG is a modern formulation, both functional and non-perturbative, of
the Wilsonian \RG, which turned out to be powerful to compute the
properties of strongly correlated systems in 
high-energy physics, condensed matter and statistical physics
\cite{Berges02,Kopietz10,Delamotte12}.  It is only recently that
turbulence has been revisited using functional and non-perturbative approaches
\cite{Tomassini97,Monasterio12,Canet16,Tarpin18,Tarpin19}. 
Exact results were obtained in this framework for the time dependence of 
 multi-point correlation functions of the stationary turbulent flow in the limit of large
  wave-numbers \cite{Tarpin18,Tarpin19}.
 These results hence  pave the way towards a deeper understanding of the fundamental properties of turbulence.

The starting point is the forced \NS equation
for incompressible flows   
\begin{equation}
  \p_t \vv+ \vv \cdot \vnabla \vv=-\frac 1\rho 
\vnabla p +\nu \nabla^2 \vv +\vf \, , \quad \vec\nabla \cdot \vv = 0 \, 
\label{eq:ns}
\end{equation}
 where $\nu$ is the kinematic viscosity and 
$\rho$ the density of the fluid.
We are interested in universal properties of turbulence. These properties are expected 
not to depend in particular on the precise form of the forcing mechanism.
We hence consider, as common in field-theoretical approaches \cite{Adzhemyan99},
 a stochastic forcing, which has a Gaussian distribution with 
 zero mean and variance
\begin{equation}\label{variance_f}
 \langle f_i(t,\vx)f_j(t',\vx\,')\rangle= \delta(t-t')N_{L^{\text{-}1},{ij}}(|\vx-\vx\,'|) \, ,
\end{equation}
where $\langle \cdot \rangle$ denotes the ensemble average over the realizations of $f$.
 This correlator is local in time, to preserve Galilean invariance,
 and is concentrated, in Fourier space, on the inverse of the integral scale $L$\footnote{
 Let us emphasize that the precise profile chosen for $N$ is not important as it 
 does not influence the universal properties of the flow, 
 as was shown in \cite{Tomassini97}. It can also be chosen diagonal in component space,
 without loss of generality because of incompressibility \cite{Canet16}.
    Moreover,  although one may argue that a forcing uncorrelated in time is not realistic physically,
    it was shown that it plays no role for the universal properties. Indeed,
  introducing finite time correlations in (\ref{variance_f}) does not alter the universal properties, as long as these correlations
   are not too long-ranged,
   as was shown in \cite{Antonov18} for Navier-Stokes equation with a  power-law forcing and in \cite{Squizzato19} for Burgers equation
    with both short-range and power-law forcing.}.
The stochastically forced  \NS equation (\ref{eq:ns}) can then be represented as a field theory following the
 standard Martin-Siggia-Rose-Janssen-de Dominicis (MSRJD) response functional formalism \citep{Martin73,Janssen76,Dominicis76},
which by essence includes all the fluctuations.  In the \RG treatment,
the integration of these fluctuations is achieved  {\it progressively} in the wave-number space,
  and yield a \textit{renormalization flow}. 
Any $n$-point correlation function $G^{(n)}$ between velocity or pressure fields
can be simply expressed in the field theory from the cumulant generating functional and obeys an exact \RG flow equation.
However, the flow equation for $G^{(n)}$ involves $G^{(n+1)}$ and
$G^{(n+2)}$ which results in a infinite hierarchy of flow equations
that needs to be solved.

 It was realized in \cite{Canet16,Tarpin18} that these  flow equations
  can be closed exactly in the limit of large wave-numbers using the symmetries
   of the \NS field theory (or more precisely extended symmetries).
Indeed,  symmetries play a key role in field theory in general,
since  they yield exact relations between correlation functions known as 
(\textit{Ward identities}). It turns out that the \NS field theory
 possesses two extended symmetries, the time-dependent Galilean symmetry
 and a time-dependent shift symmetry recently unveiled \cite{Canet15}, which 
  can be exploited to achieve the exact closure of the hierarchy of flow equations
   at large wave-numbers. Moreover, these flow equations can be solved
analytically at the fixed point, which corresponds to the stationary
turbulent state. 
Let us now
give the result for  the (transverse part of the) velocity-velocity
correlation function $C$ defined as 
\begin{equation}
C(t,\vk) \equiv \mathrm{TF}
           \avg{\vv(t_0+t,\vx_0+\vx) \cdot \vv(t_0,\vx_0)},\\
\label{eq:TF C}
\end{equation}
$\mathrm{TF}$ denoting the Fourier transform with respect to $\vx$. 
At small time lag~$t$, $C(t,\,\vk)$ takes the form
\begin{equation}
C(t,\,\vk)= A \epsilon^{2/3} k^{-11/3} 
  \exp\big(-\gamma (\epsilon L)^{2/3} (tk)^{2} + {\cal O}(k)\big),
\label{eq:C}
\end{equation}
where  $\gamma$ and $A$ are non-universal 
constants \cite{Tarpin18}, 
$\epsilon$ is the mean energy dissipation 
rate and $L$ the integral scale. The leading 
 term in~$k^2$ in the exponential
is exact, 
whereas the factor in front of the exponential 
is not. Indeed, the exponent of the power-law   could be modified
by terms of order $\ln(k)$ in the exponential, 
entering in the indicated ${\cal O}(k)$ corrections.
The leading behaviour of the correlation function
at large wave-numbers is hence a Gaussian in the variable $tk$,
which is not the expected scaling variable.
Indeed, dimensional analysis (Kolmogorov theory) predicts a dynamical critical
exponent $z=2/3$ and thus a scaling variable $tk^z = tk^{2/3}$.
The dependence in $tk$ is thus a breaking of standard scale invariance, 
generated by intermittency, 
it induces an explicit dependence in the integral scale $L$.
 An effective value  $z=1$ for the dynamical exponent is a large correction. 
     This effect is often described  phenomenologically as the random sweeping effect \cite{Kraichnan59,Tennekes75}. 

The energy spectrum 
can be  described in the dissipative range by taking the appropriate 
$t \to 0$ limit~\cite{Canet17}. 
 One can assume that the scaling variable $tk^{2/3}$
 saturates when $t$ approaches the Kolmogorov time-scale 
$\tau_K=\sqrt{\nu/\epsilon}$ and $k$ reaches $L^{-1}$, 
since these are the two relevant scales, 
such that $tk^{2/3}\to \tau_KL^{-2/3}$. 
One thus obtains for the energy spectrum
$E(k)\equiv \lim_{t\to 0} 4\pi k^2 C(t,k)$ the expression 
\begin{equation}
E(k)= A' \,\epsilon^{2/3} (k\eta)^{-\beta}
 \exp\left(-\mu (k\eta)^{\alpha}\right),
\label{spectre}
\end{equation}
where $\beta=5/3$, $\alpha=2/3$, $\mu$ and $A'$ are positive
non-universal constants related to $\gamma$ and $A$ respectively.
This behavior is valid at large wave-numbers 
that are still controlled by the fixed point,
it should correspond to the NDR.
Indeed, in the FDR, for very small spatial scales, 	
the spectrum should be regularized by the viscosity and
should be analytical in real space, 
which means that it should decay as a pure exponential in $k$-space. 
\modif{The unusual emergence of a stretched exponential behavior   
is the manifestation of the violation of  standard scale
invariance. In  standard scale invariance, 
there would be no transitional regime
between the inertial regime and the FDR.
The value $\alpha=2/3$ arises from the discrepancy between the
 Kolmogorov dynamical exponent and the effective one.
Note that the emergence of this stretched exponential was also related in Ref.~\cite{Khurshid18} to intermittency,
  and more precisely
 to intermittent energy transfers. It is indeed shown that
  the stretched exponential is removed by filtering out in time series of the energy spectrum the large 
  intermittent bursts in energy at high wave-numbers.}

The main purpose of this article is to attest the value of the 
exponent $\alpha$ in the exponential, and thereby to test the assumption
 of saturation of the scaling variable. Since the \NPRG theory 
  does not determine the prefactor of the exponential (as explained in our comment of \eq{eq:C}),
  we   use a method which allows to estimate  $\alpha$
 irrespectively of the precise form of this prefactor, as follows.
Extracting the value of $\alpha$ from the expression of
the first logarithmic derivative of \eq{spectre}
\begin{equation}
 D_1(k\eta) = \frac{d\ln (E)}{d\ln(k\eta)} = -\beta-\mu \alpha(k\eta)^\alpha \,,
 \label{der1}
\end{equation}
would require
a three-parameter non-linear fit, which is not realistic given the typical extent of the
NDR and quality of the data in this range. \modif{Previous works evidenced the
 high sensibility of such fits to the range of fitting or the initial condition of the fit algorithm \cite{Khurshid18}.} 
 Our analysis concentrates  on the second and third logarithmic derivatives
defined as \begin{align}
 D_2(k\eta) &= \frac{d (-D_1)}{d\ln(k\eta)} = \mu \alpha^2(k\eta)^\alpha
  \label{der2}\\
D_3(k\eta) &= \frac{d\ln (D_2)}{d\ln(k\eta)} = \alpha
  \label{der3} \, .
 \end{align}
In a log-log scale, the curve of $D_2$ in the dissipative range is expected 
to be a line with slope $\alpha$, 
while the curve of $D_3$ is expected to exhibit a plateau in this range of
height $\alpha$.
Beyond reducing the number of fit parameters, the main
advantage of 
$D_2$ and $D_3$ 
is to provide a method to eliminate the contributions from sub-leading 
terms in the exponential of~\Eq{eq:C}.
\modif{Let us note that previous works predicted the existence of an energy pileup, called the bottleneck effect,
 leading to a bump of the compensated spectrum in the transitional regime
 between the IR and the NDR  \cite{Falkovich94c,Lohse95}. At the theoretical level, this effect 
  modifies the prefactor of the exponential, which we do not try to resolve here. 
  The numerical and experimental spectra we have analyzed do not show
   any noticeable bottleneck effect.
   However, it plays no role in the following analysis since
    it is eliminated in the higher derivatives $D_2$ and $D_3$ studied here.  }
In the following, we first analyze the spectra obtained from \DNS in the
dissipative range, and then present the analysis of the experimental data.

\section{Numerical data}
\label{sec:num}

\subsection{Numerical simulations}

We obtain the numerical energy spectra  from \DNS of the NS equation  under fully random 
large scale forcing and in isotropic and homogeneous conditions. 
The turbulent flow is simulated in a cubic periodic domain of size $2\pi$ with the same resolution in each of the coordinate directions. The Navier-Stokes equation is solved with the use of a pseudospectral parallelized code; time advancement is implemented with the second-order Runge-Kutta scheme \cite{Lagaert14}. 
The simulation parameters are determined by the
   \Relam and the value of $k_{\rm max} \eta$, where $k_{\rm max}$ is the maximal wave-number resolved in the spectrum.
  We perform a sequence of simulations with fixed \Relam and
increasing grid resolution. 
   In the first run the computational grid size is chosen to ensure at least $k_{\rm max} \eta = 1.5$ which
    is commonly accepted as an adequate spatial resolution for \DNS. In the following runs the solution 
    obtained in previous steps is transferred to a finer computational grid with resolution 
    $N \rightarrow 2N$, while the $\Relam$ and the forcing scales are unchanged. 
     By doing so, we increase twofold the value of $k_{\rm max} \eta$ and we therefore get access to
smaller scales in the dissipative range of the turbulence spectrum. The values of the Taylor Reynolds number, 
    size of the computational grid and associated value of $k_{\rm max} \eta$ used in the simulations are
     summarized in \tref{tab:keta}. 
\begin{table}[h]
	\begin{tabular}{|c|c|c|c|c|c|}
		\hline
		 \diagbox{\quad N}{$Re_{\lambda}$}  & \quad60\quad\quad &\quad90\quad\quad &\quad160\quad\quad & \quad240\quad\quad \\
		\hline
		256 & 3.0 & 1.5 & - & - \\
		512 & 6.0 & 3.0 & 1.5 & - \\
		1024 & 12.0 & 6.0 & 3.0 & 1.5 \\
		2048 & - & 12.0 & 6.0 & 3.0 \\
		\hline
	\end{tabular}
	\caption{Parameters of the simulations: 
	 Maximal wave-number $k_{\rm max} \eta$ as a function of  the Taylor microscale 
	 Reynolds number \Relam and the grid resolution $N$.} 
	\label{tab:keta}
\end{table}

\begin{figure}[ht]
\begin{center}
\includegraphics[width=8cm]{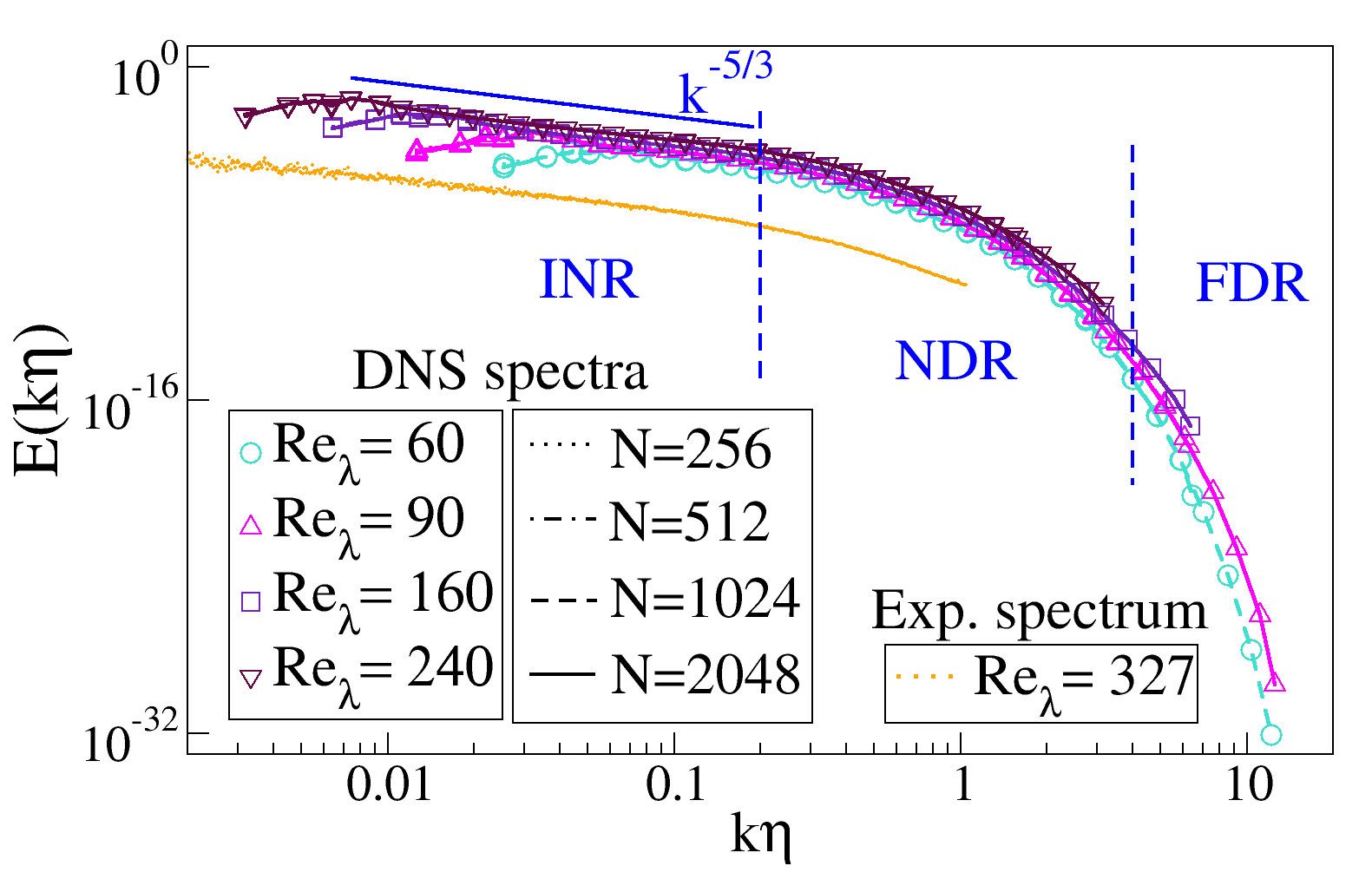}
\caption{\label{spectre-DNS} Spectra obtained from \DNS, for the set of \Relam and resolutions $N$ given in \tref{tab:keta}.
In all the figures, INR, NDR and FDR stand for inertial range, near-dissipative range and far-dissipative range respectively.}
\end{center}
\end{figure}

The spectra obtained  for each $\Relam$ and resolution, averaged over space and  time  once the 
stationary state is reached,  are displayed in \fref{spectre-DNS},
 They exhibit an inertial range extending from about half a decade at $\Relam=60$ to about a decade and a half at $\Relam=240$.
  In this range, the spectra decay as power-laws with an exponent $K$ depending on the \Relam. These values are represented on \fref{K-Re},
   together with the experimental results.  
  Both results are in agreement. Clearly, the determination of the exponent $K$ from the DNS spectra is not very precise,
 as the \Relam is limited, compared to standard \DNS
  with $k_{\rm max} \eta\simeq 1.5$.
   Let us recall that in this work we have chosen to favour the resolution over the extend of the inertial range, 
 in order to increase the  accuracy on the determination of $\alpha$.
 However, compared to previous works focused on the dissipation range, we choose a somewhat
  lower resolution (but still sufficient to resolve the NDR) while reaching a higher $\Relam$ \cite{Khurshid18}.

\begin{figure}[ht]
\begin{center}
\includegraphics[width=8cm]{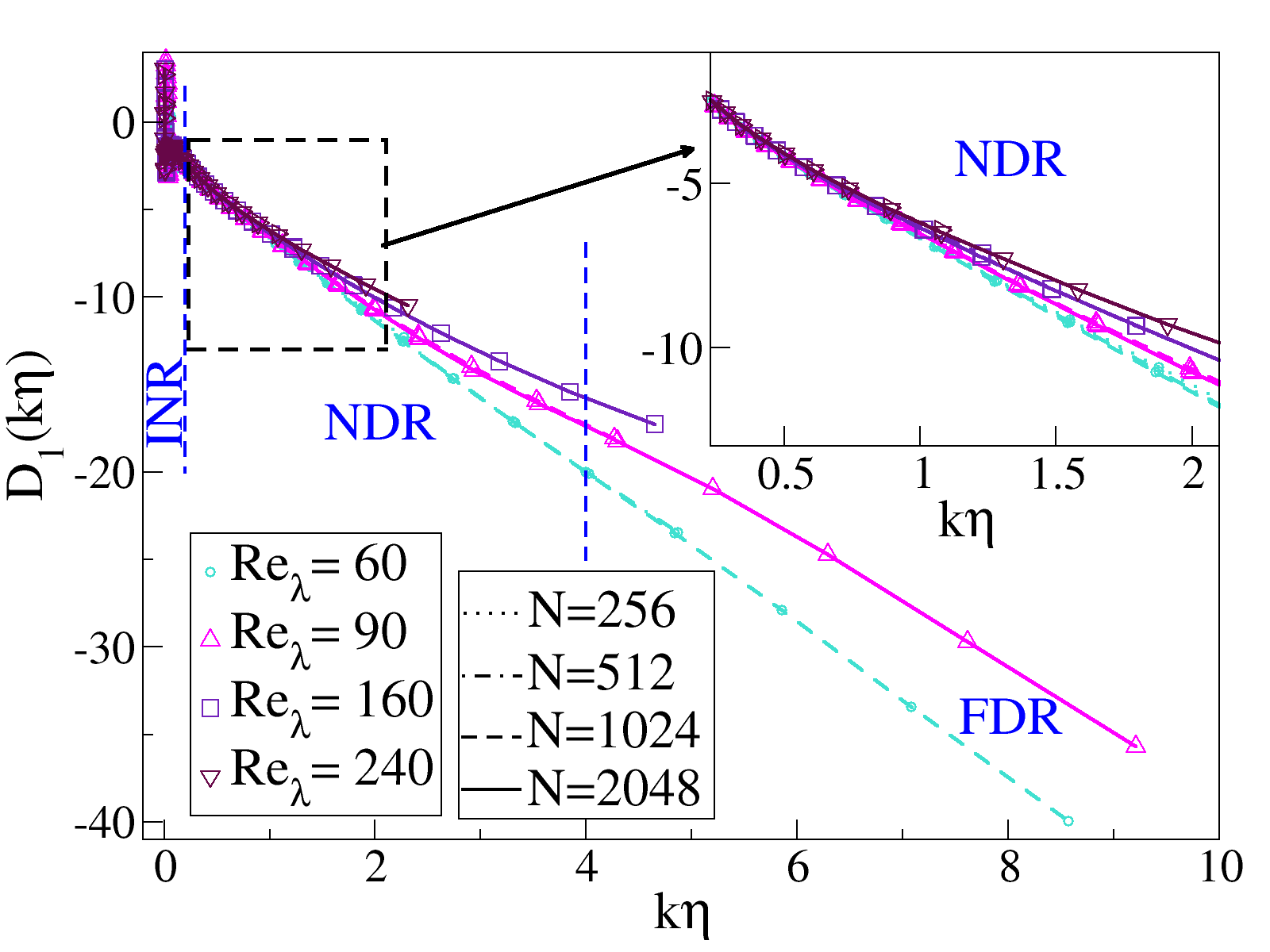}
\caption{\label{spectre-D1} First logarithmic derivative $D_1$  of the spectra of \fref{spectre-DNS}, with the inset showing a zoom in the near-dissipative range.
{These results are  
 very similar to those obtained in previous works \cite{Martinez97,Ishihara05,Khurshid18}. 
 In particular, $D_1$ is not linear in the NDR, hence it is not compatible in this range
  with a pure exponential behaviour ($\alpha=1$).}
}
\end{center}
\end{figure}

The first logarithmic derivatives $D_1$, defined by \eq{der1}, of the DNS spectra is shown in \fref{spectre-D1}.
Their behavior is  very similar to the ones 
  obtained in previous works \cite{Martinez97,Ishihara05,Khurshid18}. In particular, one distinguishes two 
  qualitatively different regimes:
   the NDR up to $k\eta\approx 4$, and the FDR extending 
   beyond this value.
    In the NDR, the $D_1$ curves exhibit a slight convex curvature, more visible as the $\Relam$  increases, which 
     indicates that $\alpha<1$ in this 
    region.   In the FDR,   the  curves of $D_1$ appear reasonably linear in log-log scale, 
    indicating a pure exponential decay $\alpha\simeq 1$. 
   Moreover, the curves do not collapse in this range, suggesting that this regime is not universal.
 These observations are in good agreement with previous studies, \modif{and lead in particular to the two-exponential model
  of Ref.~\cite{Khurshid18}, one with $\alpha<1$ dominant in the NDR, and the other one with $\alpha=1$ in the FDR.}


\subsection{Analysis of the numerical spectra in the near-dissipative range}

 In order to push further the previous observations, 
 and to obtain a precise determination of the exponent $\alpha$ in the NDR,
\begin{figure}
\begin{center}
\includegraphics[width=8cm]{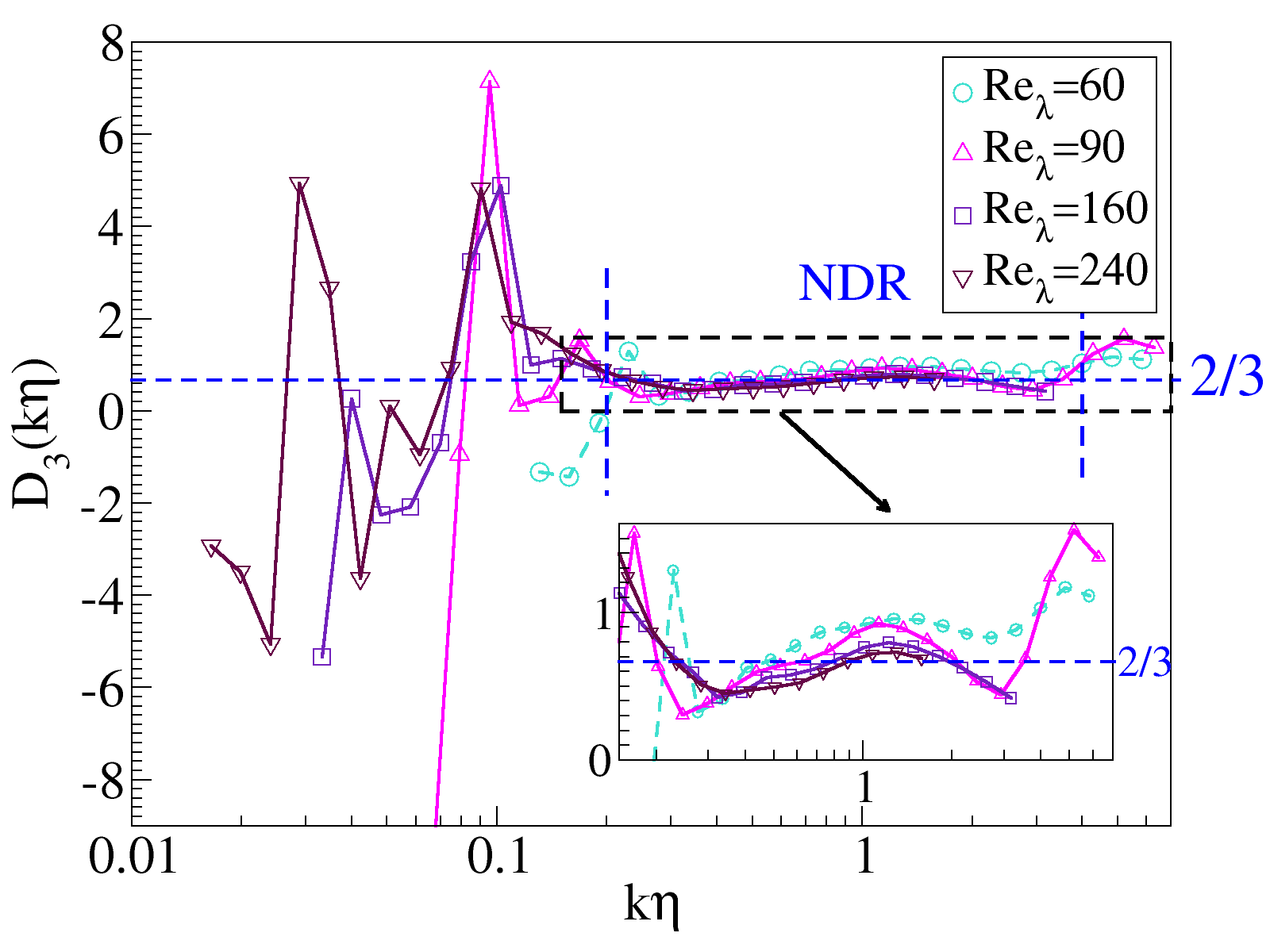}
\caption{\label{deriv-3} Third logarithmic derivative, defined by  \eq{der3}, computed for the \DNS spectra of \fref{spectre-DNS}, with the inset showing a zoom in the near-dissipative range.
Only the curves for the highest resolution for each $\Relam$ are shown for clarity. They all show a plateau in the NDR, whose value is  $\alpha$, which approaches 2/3 as the \Relam
 increases.}
\end{center}
\end{figure}
we compute the second and third logarithmic derivatives defined by \eq{der2} and \eq{der3} for the \DNS spectra
 of  \fref{spectre-DNS}. 
Our numerical data are smooth enough to allow for
 the numerical evaluation of the successive derivatives by finite differences. 
 
The result for the third derivative is displayed on \fref{deriv-3}, for the different $\Relam$'s.
 The results for the second derivative  show very  similar features.  
 The $D_3$ curves clearly exhibit  a plateau for values of $k\eta$ in the NDR, in agreement with the theoretical
  expression \eq{der3}.
  The value of this plateau is smaller than 1, and appears to be very close to the NPRG prediction $\alpha=2/3$. 
   The inset is a magnification of the plateau, which shows that $\alpha$ is close to one at small \Relam and approaches the value  $\alpha=2/3$  
   as the $\Relam$ grows. \modif{This is in excellent agreement with Ref. \cite{Khurshid18}, where they find a value of $\alpha$ in the NDR
   decreasing as $\Relam$ increases, down to $\alpha\simeq 0.78$ for their highest $\Relam=89$ 
   (we find $\alpha\simeq 0.77$ for $\Relam=90$, see \fref{Re_alpha}).}
   
   The average value of $\alpha$ on the plateau is represented in \fref{Re_alpha} as a function of the \Relam, together with the experimental results.
 Both results are in agreement.
The extent of the NDR seems to remain $0.2\lesssim k\eta\lesssim 4$
 independently of the $\Relam$.
   Interestingly, one observes that, for the spectra at lower \Relam, $D_3$ departs from the plateau at 
    higher wave-numbers -- in the FDR, towards a value which we anticipated as {~$1$}, which would signal 
    the setting of the regular (simple exponential) behavior. 
    However, the expected  value $\alpha=1$ cannot be quite reached in 
    the \DNS because of the truncation of wave-numbers at a finite $k_{\rm max}$.
    
  The analysis of the numerical spectra is thus in quantitative agreement with the theoretical prediction
  $\alpha=2/3$, and provides useful indications to guide the analysis of the experimental data.

\section{Experimental data}
\label{sec:exp}

\subsection{Data pre-processing}

Let us describe the experimental data of grid turbulence
acquired in the S1MA wind tunnel at Modane  (ESWIRP project). The data
are made of velocity time-series recorded by four hot wires at a
frequency of 250 kHz, with durations ranging from 3 to 10 minutes
each. The turbulent flow source blows air at speed ranging from 20m.s$^{-1}$ to 45m.s$^{-1}$, resulting in a variety of values for
$\Relam$. The  distance of the hot wires from the grid ranges from 7.9m to
23.16m \cite{Bourgoin17}.  We analyze all the recordings using a systematic
procedure described below, focusing on the energy spectra.

We divide the recordings  into samples of  equal time duration $\Delta t=20$s (respectively $\Delta t=60$s).
 This provides us with a total number of 3840  (respectively 1176) samples, which we analyze separately.
We use the Taylor hypothesis to express the velocity as a function of the longitudinal space coordinate.
 For each sample, we 
compute the third-order structure function
 $S_3(\ell)=\langle (\delta u_\ell)^3 \rangle$ where $\delta u_\ell$ is the longitudinal velocity increment
  $\delta u_\ell =\big( \vec u(\vec r+\vec \ell) - \vec u(\vec r)\big)\cdot \vec \ell/|\vec \ell|$.
   We determine the corresponding   average injection/dissipation rate per unit mass 
 $\epsilon$ from the plateau value of $S_3$ using the four-fifth law:  $\epsilon =-\frac 5 4 S_3(\ell)/\ell$
 in the inertial range.
 We then compute the associated Kolmogorov scale 
  as $\eta=\nu^{3/4}\epsilon^{-1/4}$, 
   with $\nu=1.5 \times10^{-5} \mathrm{m^2.s^{-1}}$ the air kinematic viscosity.
    The Taylor microscale Reynolds number for each sample  is  deduced using the isotropic relation
 $\mathrm{Re}_\lambda=u_{\rm rms}\sqrt{15/\nu\epsilon}$.
 Let us note that the determination of $\epsilon$ through the four-fifth law is not very precise, and thus it entails some error on the \Relam. However,
 we also used another determination of $\epsilon$, through the  position of the peak of the dissipative spectrum $k^2 E(k)$,
  and we checked that this does not change the different distributions shown here, nor the final value of $\alpha$.

 We compute the kinetic energy spectrum  of each sample by
performing a discrete Fourier transform (FFT algorithm) and rescale
it as a function of the  dimensionless wave-number {$k\eta$}. 
(A typical spectrum is displayed at the top of the \fref{method1}.) 
We smoothen the spectra out using a
regular binning of 20,000 bins in {$\ln(k\eta)$} scale.
  In the experimental spectra, 
  {$k\eta$} typically ranges from $10^{-3}$ to $3.5$.
  The quality and the extent in wave-number of the spectra vary quite substantially  between the samples. The variability of the measurement quality comes from the fact that the wind tunnel is an open facility in which dust and pollen can enter and affect the hot wires. 
 Because of this issue, some values of $\epsilon$ are questionable, which explains why some values of $\Relam$ are out of the range $200-600$.
 However, all of them appear to increase
as functions of {$k$} beyond a wave-number {$k_M$}, which we interpret  as a contamination by the small-scale response of the hot wires.
 The value of {$k_M$} depends on the spectra, but is found to be typically {$k_M\eta\simeq3$}.
  In the following, we restrict the analysis   to wave-numbers below  {$k_M$}.
  \modif{Note that we do not observe any noticeable bottleneck effect on the experimental
  spectra analyzed. As this effect is expected to attenuate when the $\Relam$ increases \cite{Donzis10}, 
  it would be difficult to detect here, moreover as already explained, 
  it alters the prefactor of the exponential and disappears in the higher
   derivatives considered in the present work.}
\begin{figure}[t]
\begin{center}
\includegraphics[width=8.6cm]{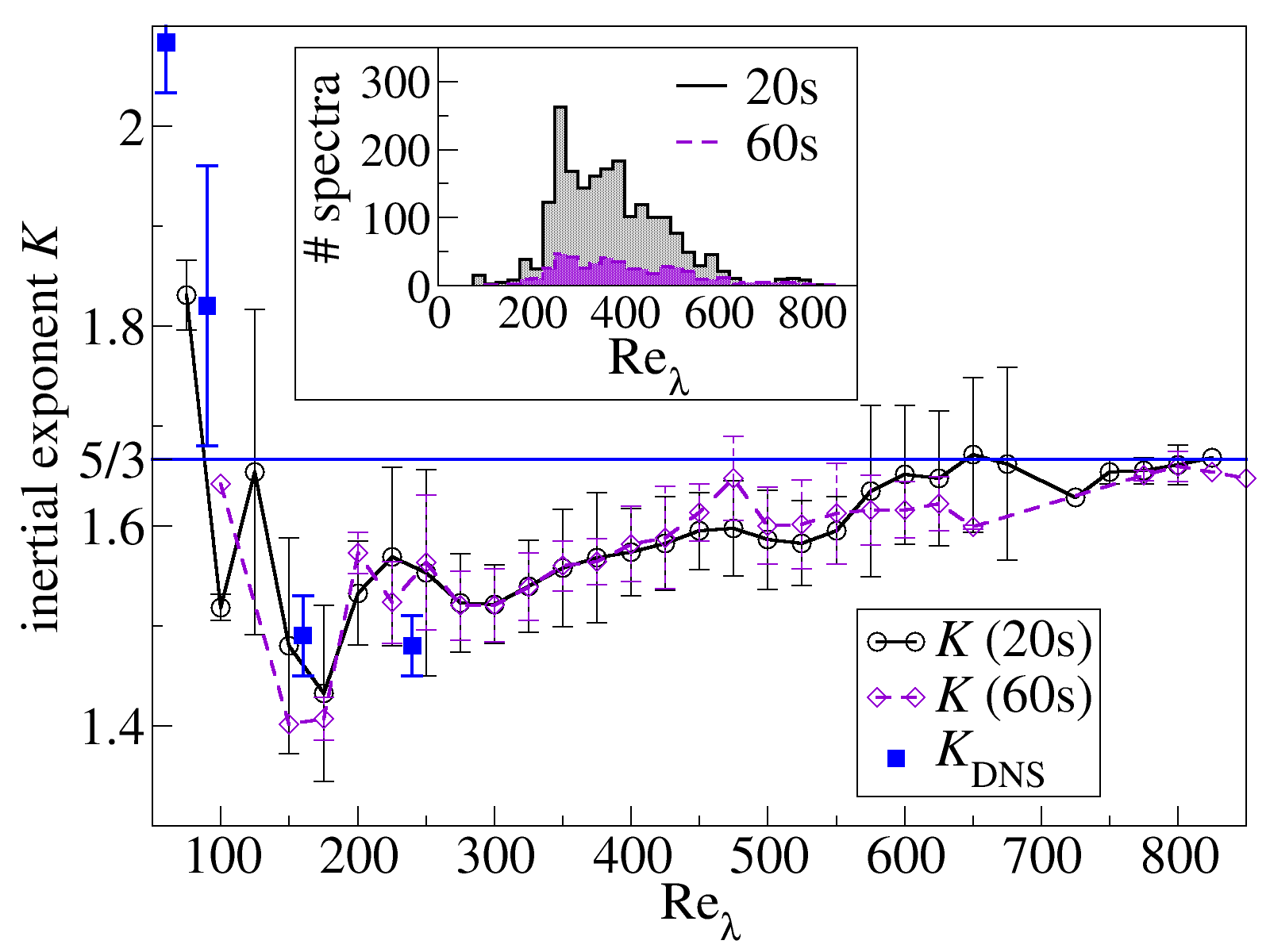}
\caption{\label{K-Re} Exponent $K$ of the power-law in the inertial range, from \DNS and from experiments for both sample durations $\Delta t=20$s and $\Delta t=60$s,  
as a function of the Taylor microscale Reynolds number.
 For the experimental points, the symbols represent the mean values of  $K$ on each $\Relam$ bin, and the error bars 
the standard deviation within each  bin. The number of spectra in each bin is not equal, it is given
 by the distribution shown in the inset. }
\end{center}
\end{figure}

\subsection{Selection of spectra}

In order to assess the quality of each sample and to remove low-quality spectra,
 we operate a data-driven selection, based on the value of the inertial range exponent $K$
  and on the extent of the dissipative range, whose determinations are detailed in  
   \sref{sec:inertial} and \sref{sec:dissipative} respectively.
We first eliminate for both sample durations $\Delta t=20$s or $\Delta t=60$s
   the spectra with an exponent $K$ differing  from the K41 value by more than $20\%$.
   This procedure typically removes  very noisy spectra presenting several unphysical large peaks.
   After this first filtering, there remain
  3486 {(respectively 1080)} exploitable spectra for $\Delta t=20$s {(respectively $\Delta t=60$s)}.
    We then proceed to a second selection by retaining only the spectra with a large enough dissipative
     range of width $\Delta (k\eta) > 0.2$. This second selection removes spectra
      which are   not well resolved in the dissipative range, probably affected by the noise 
      from the sensor.
      This finally leaves us with 1641, respectively
      590 spectra, for $\Delta t=20$s and $\Delta t=60$s. We only further analyze and discuss these selected spectra
       in the following.

\subsection{Inertial range}
\label{sec:inertial}

 In this section, we briefly describe the analysis of the spectra in the inertial range.
We determine the exponent $K$ of the power-law decay of the energy spectrum $E(k\eta)\sim (k\eta)^{-K}$ in this range, to be compared
 with the expected K41 value $K\simeq 5/3$. We do not denote it $\beta$ as in \eq{spectre} since we do not assume the 
 two exponents to be equal in the inertial and dissipative ranges.
 We use two methods to estimate $K$:   a direct fit of the log-log spectrum -- which is expected to be a line in 
 the inertial range,  and a fit of the first logarithmic derivative $D_1$ -- which is expected to present a plateau of 
 value $-K$ in the corresponding range. 
 Throughout this work, we only use fitting functions which are affine (either lines or  constants). 
 The result of a fitting procedure is in general sensitive to the precise fitting interval chosen, 
 and to the chosen precision criterion.
  In order to reduce as much as possible these effects, we 
   devise an optimized algorithm, described in \aref{app:num}, 
   whose goal is to search for the largest interval with the minimal error where an affine behavior is present.
 For each spectrum, we apply this fitting algorithm to both the log-log spectrum and the corresponding $D_1$,
  and compare these two results,   If the relative difference in the obtained $K$ is less than 2\%,
   we take the average value. Otherwise we keep the estimate for $K$ which corresponds to the largest interval.

\begin{figure}[t]
\begin{center}
\includegraphics[width=8.6cm]{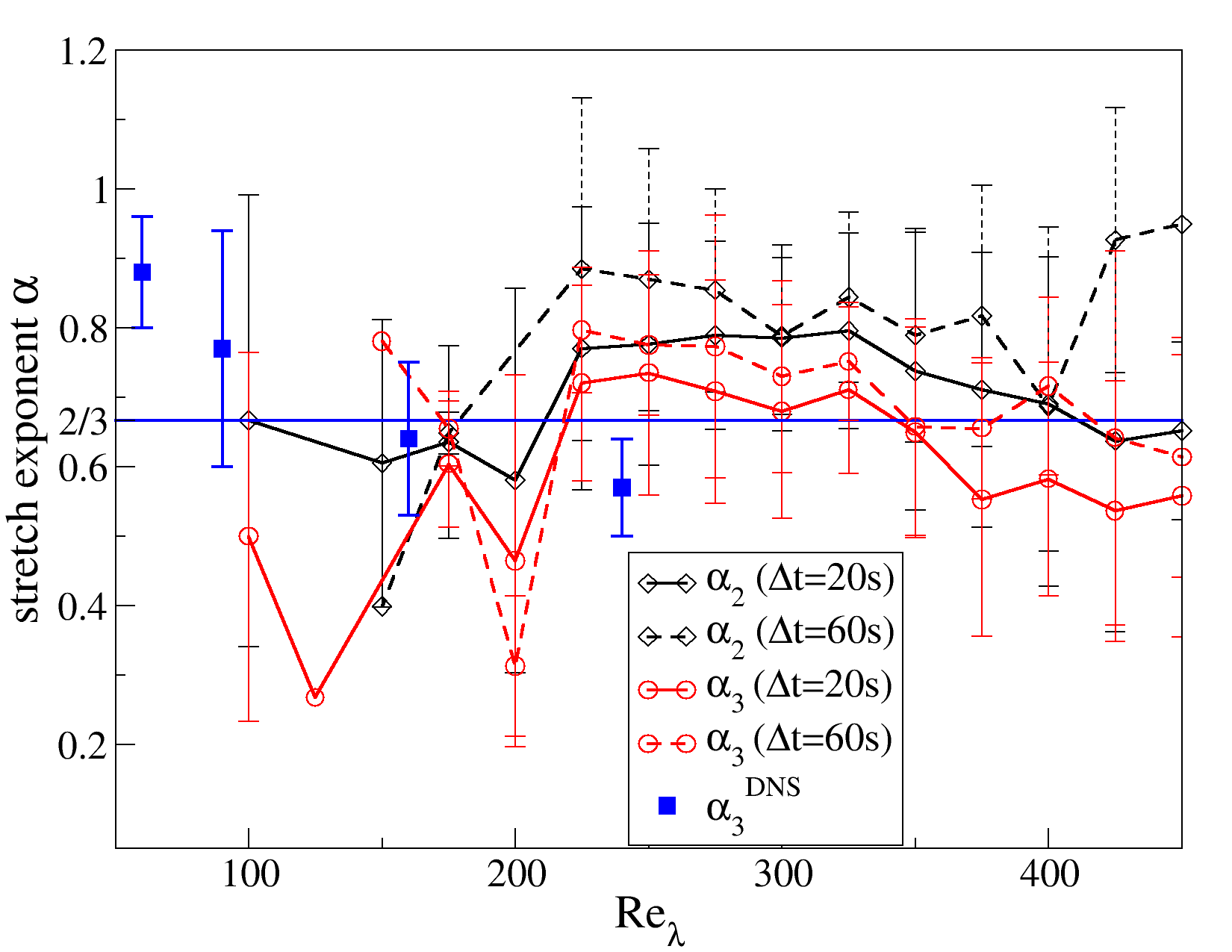}
\caption{\label{Re_alpha} Stretch exponent $\alpha$ as a function of the Taylor microscale Reynolds number from \DNS and from experiments: $\alpha_2$  (black lines with diamonds) computed from $D_2$ and  $\alpha_3$ (red lines with triangles) computed from $D_3$,
 for experimental spectra corresponding to both duration $\Delta t =20s$ (plain lines) and  $\Delta t =60s$ (dashed lines), and $\alpha_3$ (blue squares) from \DNS.
For the experimental points, the symbols indicate the mean values of  $\alpha$ for each $\Relam$ bin, and the error bars 
 the standard deviation within each  bin. The number of spectra in each bin is not equal, the distribution of $\Relam$ for each set of spectra is represented in  \fref{dist-Re}.} 
\end{center}
\end{figure}
\begin{figure}[t]
\begin{center}
\includegraphics[width=8.6cm]{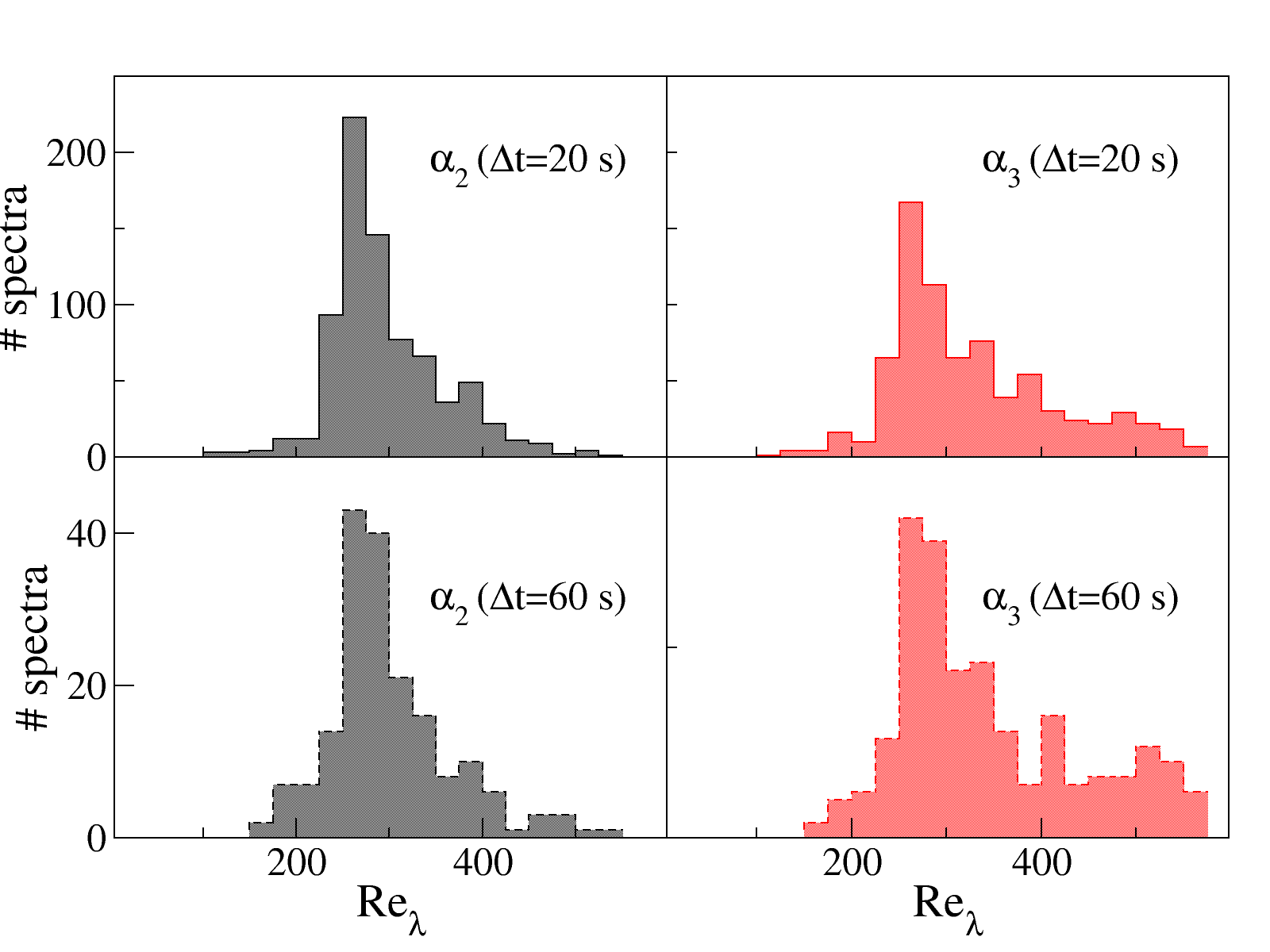}
\caption{\label{dist-Re} Distribution of the $\Relam$ corresponding to each set of data displayed 
in \fref{Re_alpha}, for the two sample durations $\Delta t=20s$ and $\Delta t=60s$, and for the two methods $D_2$ and $D_3$. } 
\end{center}
\end{figure}

  We present in \fref{K-Re} the results for the exponent $K$ as a function of the \Relam,
   for the two sample durations  $\Delta t=20$s and $\Delta t=60$s.
 The experimental data 
correspond to \Relam distributed between 25 and 1150, among which more than $90\%$ lie between
 150 and 650. Their distribution is represented in the inset of \fref{K-Re}, using bins of width  $\Delta \Relam =25$,
  and computing within each bin
  the average and the standard deviation of the decay exponent $K$.
   The error bars in \fref{K-Re} represent the standard deviation 
   within each bin, irrespective of the total number of spectra they contain.  Note that the weight of the
    different points is thus very different: for instance  the first points at low $\Relam$ represent very few spectra.
We observe that the value of the inertial exponent $K$ grows with the $\Relam$, to reach a value close to {$K=5/3$} for
 $\Relam\gtrsim 600$. The results for both durations are in close agreement.   However, let us emphasize
 that this determination is not accurate enough to detect the corrections due to intermittency, which are expected to
  increase by about 5\% the K41 value $K=5/3$. The error bars of the points around $\Relam =600$
 obtained for the duration $20$s are typically representative since they correspond to a large number of spectra, and they
 clearly exceed the 5\% level of tolerance needed to estimate intermittency corrections on the inertial range exponent.

\subsection{Near-dissipative range and stretch exponent}
\label{sec:dissipative}

In this section, we turn to the analysis of the near-dissipative range.
 As for the \DNS data, we rely for the experimental data on two independent 
 estimations of the stretch exponent $\alpha$: 
  one from the second logarithmic derivative
 $D_2$ given by \eq{der2} and one from the third one $D_3$ given by \eq{der3}.
 The experimental data are much less smooth than the numerical ones, such that performing  
  simple finite differences to compute numerical derivatives 
   generates
  a lot of noise, which would spoil the accuracy of this procedure. For this reason, we resort to a 
  more reliable 
  computation of the derivatives, described in \aref{app:num}. 
  We obtain two determinations of $\alpha$ denoted  $\alpha_2$ and $\alpha_3$,
   from the optimized linear fitting of $D_2$ in log-log scale, and from applying to $D_3$
    the same optimized algorithm to search for a plateau respectively.
    We only retain estimations of $\alpha$ for which the error of the associated fit is less than $Q=10^{-1}$
     (see \aref{app:num} for the definition of $Q$).

    We present on  \fref{Re_alpha} the results obtained for $\alpha_2$ and $\alpha_3$  as  functions of  the $\Relam$,
for both sample durations $\Delta t =20$s and $\Delta t =60$s.
The distribution of  $\Relam$ corresponding to the four sets of spectra are displayed in \fref{dist-Re}. 
The large majority of spectra presents a $\Relam$  again lying in the range $200\lesssim \Relam\lesssim 500$.
 The number of spectra with a $\Relam$ outside this range is negligible.
    Contrary to the inertial exponent $K$,
     we do not observe for $\alpha$ a clear dependence on the $\Relam$ (at least not within the present level of accuracy),
      but rather a  constant value (up to fluctuations) smaller than 1. 
      The error bars on \fref{Re_alpha} represents the standard deviation within each $\Relam$ bin, and do not reflect
       the fact that some bins (at lower or larger $\Relam$) contain very few spectra.
       Moreover, 
      one notices that the estimation of $\alpha_2$ appears systematically  larger {than} the one from
        $\alpha_3$. The determination of $\alpha_2$ involves a two-parameter fit, such that the resulting 
         value for $\alpha_2$ is expected to be less precise than the one of $\alpha_3$,
          but we have no interpretation for the seemingly systematic  over-estimation.

\begin{figure}
\begin{center}
\includegraphics[width=8.6cm]{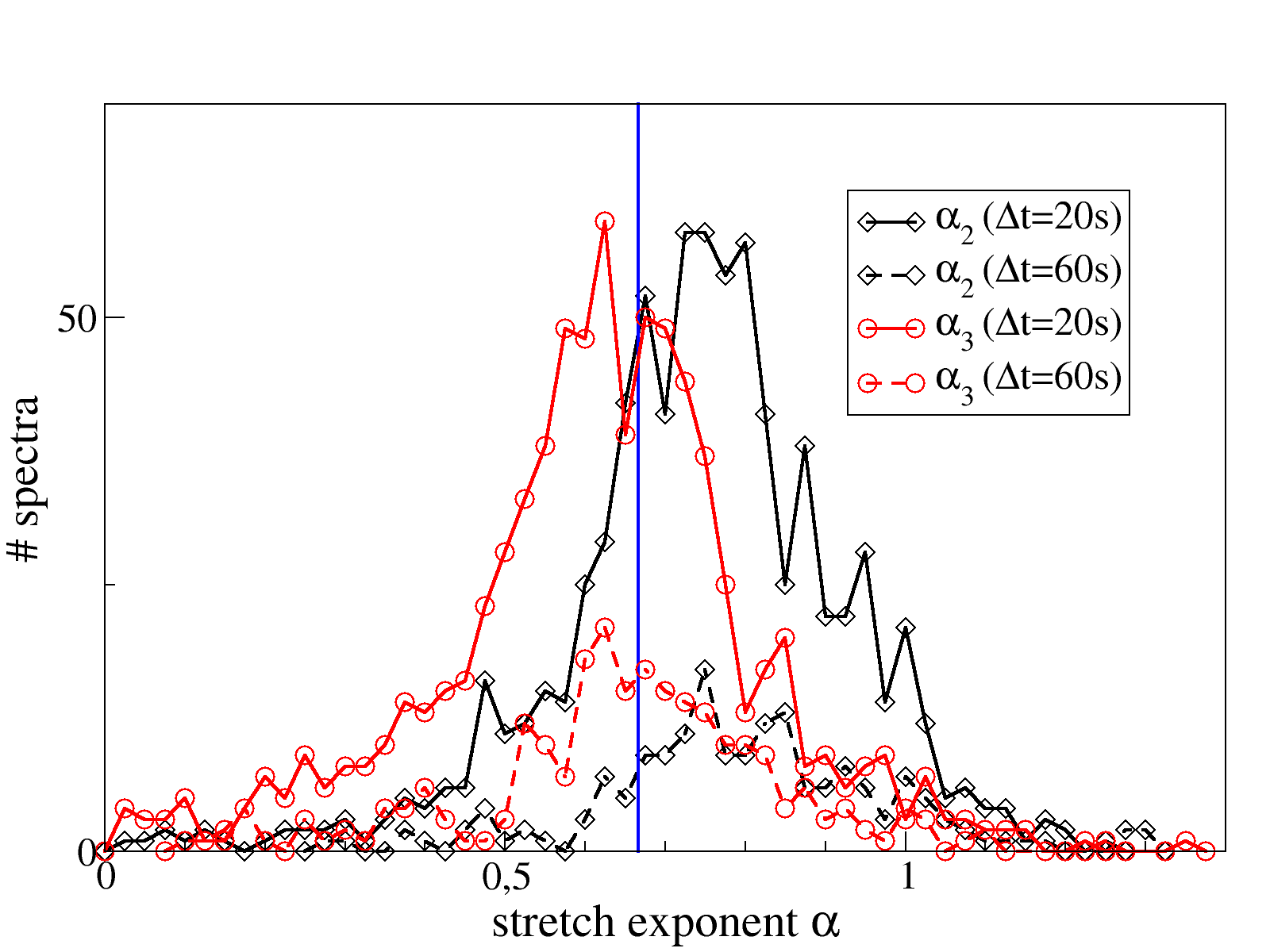}
\end{center}
\caption{\label{distrib} Distribution of the stretch exponent $\alpha$: $\alpha_2$  (black lines with diamonds) 
computed from $D_2$ and  $\alpha_3$ (red lines with triangles) computed from $D_3$,
 for spectra corresponding to both durations $\Delta t =20s$ (plain lines) and  $\Delta t =60s$ (dashed lines).
  The mean value and standard deviation of these distributions are reported in Table \ref{tab:alpha}.}
\end{figure}

\begin{figure}
\begin{center}
\includegraphics[width=8.6cm]{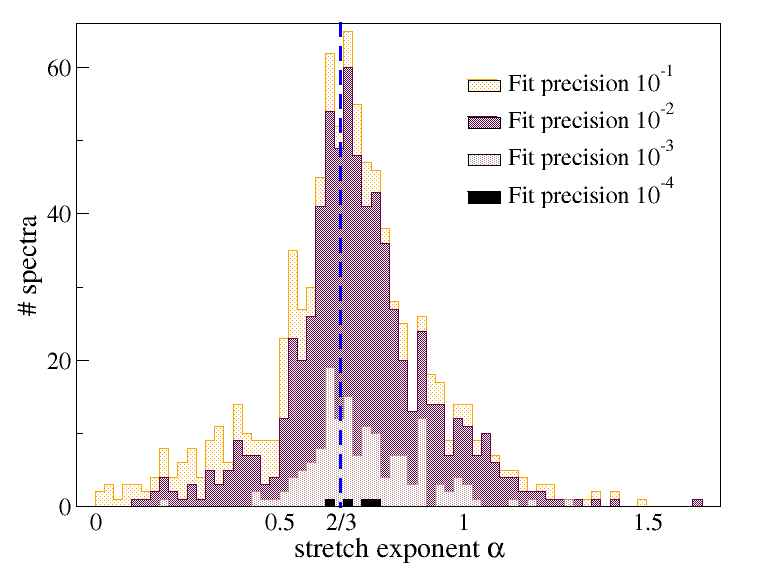}
\end{center}
\caption{\label{distrib_prec} Distribution of the stretch exponent $\alpha_3$ for $\Delta=60$s
 and for different precision criteria $Q$ from $10^{-1}$ to $10^{-4}$. }
\end{figure}

Since we did not find any sizeable dependence of the stretch exponent $\alpha$ on the $\Relam$,
 we rather  concentrate in the following on the  distributions of the  values of $\alpha$.
 We present in \fref{distrib} these distributions  for the same data as the one used in \fref{Re_alpha}, 
 that is both $\alpha_2$ and $\alpha_3$ and 
 for both sample durations. One notices again that the distributions for $\alpha_2$
 are slightly shifted upward with respect to the ones for $\alpha_3$. The  averages  and standard deviations
  of all the distributions are gathered in \tref{tab:alpha}. 
 The distributions  presented in \fref{distrib} correspond to an error tolerance
  of $Q=10^{-2}$. We studied the influence of this precision criterion by varying
   it from $10^{-1}$ to $10^{-4}$. The corresponding distributions for $\alpha_3$ are displayed in \fref{distrib_prec}.
   It is remarkable that
   the more stringent the precision criterion, the narrower the distributions around 
   the theoretical prediction $\alpha=2/3$. In fact, very few spectra pass the highest precision criterion
    $Q=10^{-4}$ for the sample duration $\Delta t=60$s, and none for  $\Delta t=20$s.
  
To give a final number for $\alpha$, we can favor the most precise determination, 
 which is $\alpha_3$. We retain the  two precision criteria $Q=10^{-2}$ and $Q=10^{-3}$, which embody a reasonable compromise
  between the level of accuracy and the level of statistics. Finally, since the spectra for both sample durations constitute independent sets of comparable
  qualities, we can consider a weighted average of these results, which
   yields 
 \begin{equation*}
 \langle\alpha_3\rangle_{10^{-2}} \simeq 0.677 \pm 0.193\quad,\quad \langle\alpha_3\rangle_{10^{-3}} \simeq 0.712 \pm 0.157\, .
 \end{equation*}
 Both estimates are very close, since their one-$\sigma$ intervals are respectively
 $\alpha\in[0.48:0.87]$ and $\alpha\in[0.55:0.87]$. We hence choose as a representative final result
 \begin{equation*}
 \langle\alpha\rangle \simeq 0.68 \pm 0.19 \, .
 \end{equation*}
 The present analysis indicates that  values $\alpha<0.4$ or $\alpha>0.9$ can be reasonably excluded.
 
\begin{table}
\begin{tabular}{|p{1cm}|p{1cm}|p{1cm} |cll|}
\hline
$\alpha_i$ & $\Delta t$ & $Q$ & \#  &  \;\;$\langle\alpha\rangle$ & \;\;$\sigma$ \\
\hline
\multirow{6}{3cm}{$\alpha_2$} & \multirow{3}{1cm}{20s}
 & $10^{-1}$ &\; \;\;792\;\;\; & 0.754 \;\;& 0.181\;\; \\
&& $10^{-2}$ & 785 & 0.757 & 0.176 \\
&& $10^{-3}$ & 685 & 0.782 & 0.151 \\
&& $10^{-4}$ & 282 & 0.756 & 0.126 \\
\cline{2-6}
& \multirow{3}{1cm}{60s}
 & $10^{-1}$ & 178 & 0.805 & 0.188 \\
&& $10^{-2}$ & 175 & 0.812 & 0.178 \\
&& $10^{-3}$ & 162 & 0.833 & 0.159 \\
&& $10^{-4}$ & 99 & 0.829 & 0.126 \\
\hline
\multirow{5}{3cm}{$\alpha_3$} & \multirow{2}{1cm}{20s}
 & $10^{-1}$ & 2892 & 0.621 & 0.226 \\
&& $10^{-2}$ & 1798 & 0.661 & 0.194 \\
&& $10^{-3}$ & 89 & 0.678 & 0.167 \\
&& $10^{-4}$ & 0  &  --   & -- \\
\cline{2-6}
& \multirow{3}{1cm}{60s}
& $10^{-1}$ & 878 & 0.690 & 0.224 \\
&& $10^{-2}$ & 703 & 0.719 & 0.190 \\
&& $10^{-3}$ & 153 & 0.730 & 0.151 \\
&& $10^{-4}$ & 4 & 0.710 & 0.052 \\
\hline
\end{tabular}
\caption{\label{tab:alpha} Summary of the number of spectra (\#), average value ($\langle\alpha\rangle$) and standard deviation ($\sigma$) of the distributions of 
the stretch exponent $\alpha$
 obtained from either the second ($\alpha_2$) or the third ($\alpha_3$) logarithmic derivatives, 
 for spectra of both sample durations $\Delta t=20$s and $\Delta t=20$s, and for different
  precision tolerances $Q$ on the fit.}
\end{table}

\subsection{Near-dissipative range: other parameters}

\begin{figure}[t]
\begin{center}
\includegraphics[width=8.6cm]{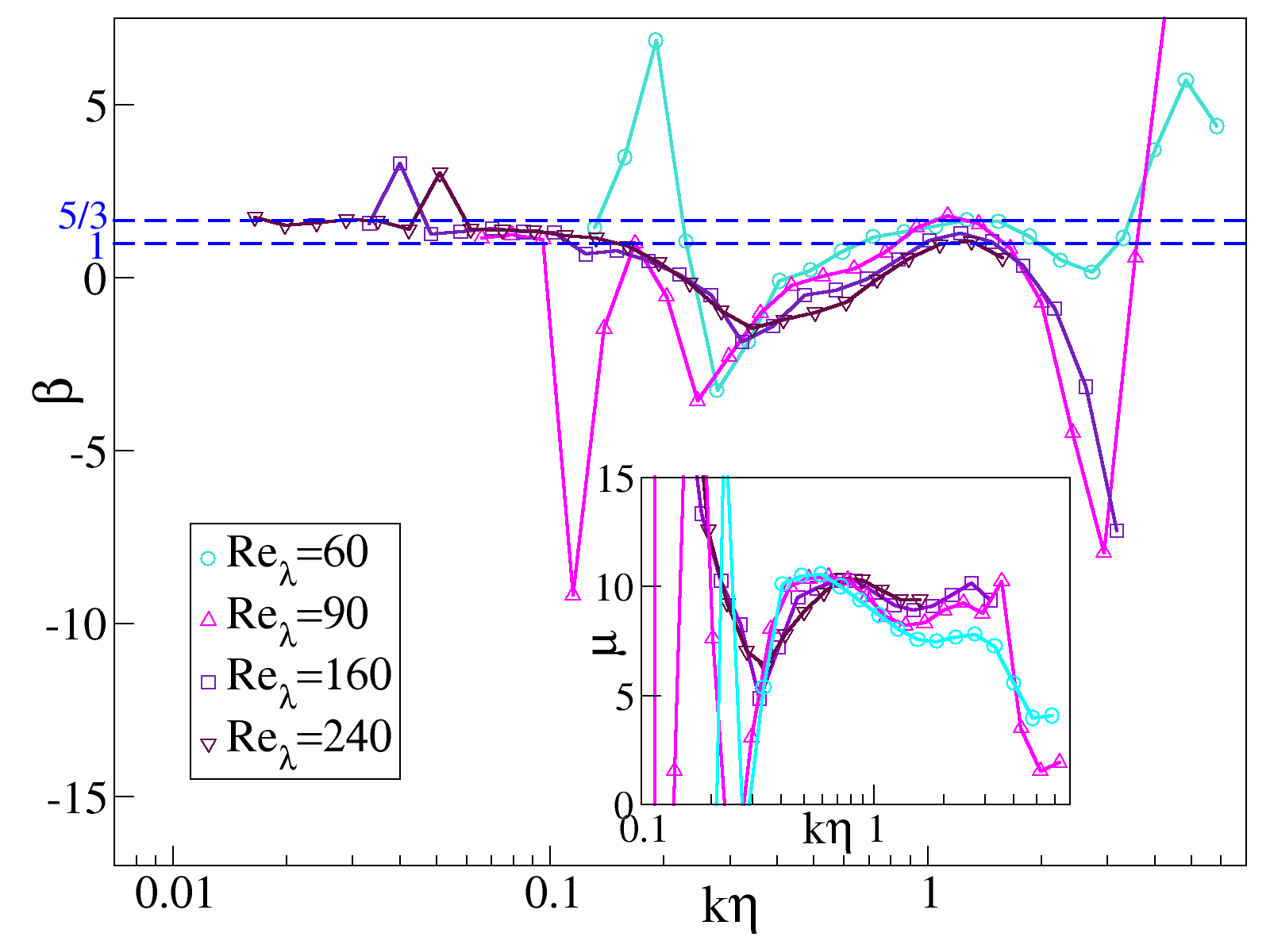}
\caption{\label{beta-mu} Parameters $\beta$ (main graph) and $\mu$ (inset) obtained for the \DNS spectra of \fref{spectre-DNS}.} 
\end{center}
\end{figure} 
For completeness, we also give in this section the results for the other parameters of \eq{spectre}:
 the multiplicative constant  $\mu$ in the exponential,
 and the exponent $\beta$ of the power-law.
The parameter $\mu$ can be obtained from the constant $\mu\alpha^2$ in the linear fit of $D_2$ in log-log scale
according to \eq{der2}. The values found for the experimental data are given in \tref{tab:betamu}. If we consider, as we do for $\alpha$,
the weighted average of both sample durations for $Q=10^{-2}$, we obtain:
 \begin{align*}
 &\langle\mu\rangle \simeq 4.45 \pm 1.31 \, ,
 \end{align*}
 which turns out to be of the same order  as values found in previous studies \cite{Debue18}, although $\mu$
  is not expected to be universal. In the numerical data, we obtain somewhat higher values, of order  10, as shown in \fref{beta-mu}.
 
 The value of $\beta$ can be computed as
\begin{equation}
  \beta= \frac{D_2(k\eta)}{D_3(k\eta)} - D_1(k\eta) \,.
\end{equation}
However, the precision on $\beta$ is very poor, and it is highly sensitive to the value found for $\alpha$.
As an indication, we find in the experimental data values close to $1$, as summarized in \tref{tab:betamu}. For instance for $Q=10^{-2}$,
we obtain from the weighted average of both samples durations:
  \begin{align*}
  &\langle\beta\rangle \simeq 0.93 \pm 0.26 \,.
 \end{align*} 
 In the \DNS, the  value obtained for $\beta$ is shown in \fref{beta-mu}. It turns out to be close to $\beta=5/3$ {\ie}
 to the inertial range value) at small $\Relam$, and to tend to a smaller value very close to $\beta=1$ as $\Relam$
   increases, in agreement with the results from the experimental data. But once again, these estimates should be taken with caution.
Given the typically small extent of the NDR, the determination of the leading behavior, through $\alpha$,
 is already challenging, such that a precise determination of the sub-leading one, through $\beta$,
  is beyond the reach of the present work.
\begin{table}
\begin{tabular}{|p{1cm}|p{1cm}|c|cc|cc|}
\hline
$\Delta t$ & $Q$ & \;\;\;\;\# \;\;\;\; & $\langle\mu\rangle$ & $\sigma_\mu$ & $\langle\beta\rangle$ & $\sigma_\beta$\\
\hline
\multirow{3}{1cm}{20s} 
& $10^{-1}$ & 750 & 4.51 & 1.26  & 0.92 & 0.26\\
& $10^{-2}$ & 747 & \;\; 4.51 \;\;& \;\; 1.26 \;\;&\;\, 0.92\;\; & \; 0.26\;\;\\
& $10^{-3}$ & 672 & 4.46 & 1.23 & 0.93 & 0.27\\
& $10^{-4}$ & 275 & 4.53 & 1.07 & 0.87 & 0.25\\
\hline
\multirow{3}{1cm}{60s} 
& $10^{-1}$ & 173 & 4.16 & 1.54 & 0.99 & 0.27\\
& $10^{-2}$ & 172 & 4.19 & 1.51 & 0.98 & 0.26\\
& $10^{-3}$ & 162 & 4.06 & 1.28 & 1.00 & 0.23\\
& $10^{-4}$ & 99 & 3.90 & 0.89 & 1.00 & 0.19\\
\hline
\end{tabular}
\caption{\label{tab:betamu}Average value and standard deviation of the distributions of the parameters $\mu$ and $\beta$
 obtained for spectra of both sample durations $\Delta t=20$s and $\Delta t=60$s, and for different
  precision tolerances $Q$ on the fit.}
\end{table}

\section{Conclusion}

In this work, we analyze the near-dissipative range of kinetic energy spectra obtained from high-resolution \DNS 
 and from experimental grid turbulence in the Modane wind tunnel in order to test the \NPRG prediction
  of a stretched exponential
  behavior $E\propto \exp(-\mu (k\eta)^\alpha)$ with exponent $\alpha=2/3$ in this range.
  We use two independent determinations, from the second and from the third
   logarithmic derivatives of the spectra, to estimate the value of $\alpha$.
   All the results, from \DNS and from experiments, and
    from the different determinations,  are consistent, and yield an estimate 
  for the stretch exponent  $\alpha \simeq 0.68 \pm 0.19$,
   in full agreement with the theoretical prediction.
  
  Let us emphasize that the typically small extent of the near-dissipative range 
  and the level of precision currently accessible in \DNS or experimental data
   do not seem sufficient to reliably determine the prefactor of the exponential, that is the
    precise form of the power-laws. Hence, the present analysis does not allow us to shed a new light 
    on this point. It could possibly be refined in the future if data with still higher
     resolution  become available. Moreover, it would be also very interesting to further test
      the NPRG predictions concerning the time dependence of two- or multi-point correlation functions,
       in \DNS or in experiments which dispose of  suitable measurement techniques.

\begin{acknowledgments}
LC and VR gratefully thank N. Wschebor for fruitful discussions and useful
suggestions.  
We acknowledge the European Union for its support and access to
the ONERA operated S1MA wind tunnel through the ESWIRP project (FP7/2007-2013
under grant agreement 227816).
This work received support from the French ANR through the project NeqFluids 
(grant ANR-18-CE92-0019). 
The simulations were performed using the high performance computing resources
from GENCI-IDRIS (grant 020611).  
GB and LC  are grateful for the support of the 
Institut Universitaire de France.

\end{acknowledgments}

\appendix

\section{Numerical procedures}
\label{app:num}

\subsection{Optimized linear fitting algorithm}

 Extracting the inertial or dissipative exponents from the experimental data
  always amounts to an affine fit (either lines  in log-log representation or constants).
  For this, we use an elementary linear
regression. However, the accuracy of the result of such a fit
strongly depends on the choice of the fitting domain $D$, which is delicate
 since there is no clear delimitation of the inertial or dissipative
  ranges. 
Let us denote by $a(D)$ and $b(D)$ the real numbers such that the standard
linear fit in the domain $D$ is the affine function $x\mapsto f(x)=a(D)x+b(D)$
and by $x_i$ a data point in the domain $D$, {\ie} $x_i\in D$, and $y_i=f(x_i)$. 

To determine the best fitting domain, we use an optimized method based on the alignment of
data points in a given domain $D$. The alignment $\lambda(D)$ is the largest
absolute deviation of a data point from the linear fit performed in this domain: 
$\lambda(D)=\max_{x_i\in D}|a(D)x_i+b(D)-y_i|$. 
Based on $\lambda(D)$, our algorithm seeks the largest domain
$D(\varepsilon)$ such that $\lambda(D)<\varepsilon$ for any given positive
number $\varepsilon$. $\varepsilon$ stands as a quality requirement of the
input data.
We estimate the quality $Q(D)$ of the result using the
standard error $Q=\sigma/\sqrt n$ where $\sigma$ is the standard deviation
and $n$ is the number of data points in the domain.
Assembling these elements, we determine the value $\varepsilon_{\rm
opt}$ that minimizes the function $Q(D(\varepsilon))$ and use
$D(\varepsilon_{\rm opt})$ as the optimal domain.
 This procedure is used to
determine the inertial exponent $K$ of the spectrum and the stretch
exponent $\alpha$ from the second logarithmic derivative $D_2(k\eta)$.  

We use the same procedure to find the optimal plateau,  {\ie} the
optimal domain where the curve has the lowest slope.
This is achieved by replacing  $Q$ by $Q'(D)=Q(D)+|a(D)|$, where $a(D)$
is as before the slope of the linear fit in the domain $D$. We use this method to
determine the inertial exponent $K$ from the first logarithmic derivative $D_1(k\eta)$
 and for  the stretch exponent $\alpha_3$ from the third logarithmic derivative $D_3(k\eta)$.
The value of $Q$ is recorded for each estimate, and is used as a precision indicator
  for the  quality of the fit. In particular, we rely on $Q$ to set up
    different precision criterion, {\it eg.} in \fref{distrib_prec} or \tref{tab:alpha}. 
The application of these procedures is illustrated in \sref{app:illust} on a typical spectrum.

\subsection{Numerical computation of derivatives for the experimental data}

As explained in \sref{sec:dissipative}, the direct computation by finite differences
 of successive derivatives of the experimental spectra is too noisy to be exploitable.
  Instead,  to smooth out the data, we use the following procedure.
   The derivative of a function at a certain point 
   is computed by implementing  a linear
fit in a window $D$ centered around this point containing a fixed number of data points (we  typically use 
windows of 50 points). The linear fit coefficient $a(D)$ is thus an estimate of  the value of
the derivative at this point. By sliding the window $D$ through all the data points, we  obtain the
 derivative of the whole function. The logarithmic 
derivatives of the type $d \ln f/ d\ln q$ are performed using the same method with a power
law fit (fitting function $q\mapsto b(D) q^{a(D)}$) and the derivatives of the type $d/d\ln
q$ with an exponential fit (fitting function $q\mapsto b(D){\mathrm e}^{a(D) q}$).
 With this procedure, the resulting derivatives of the experimental spectra are smooth enough
  to serve for further analysis.

\subsection{Illustration of the determination of the exponents $K$ and $\alpha$}
\label{app:illust}

  \begin{figure}[h]
\begin{center}
\includegraphics[width=6.cm]{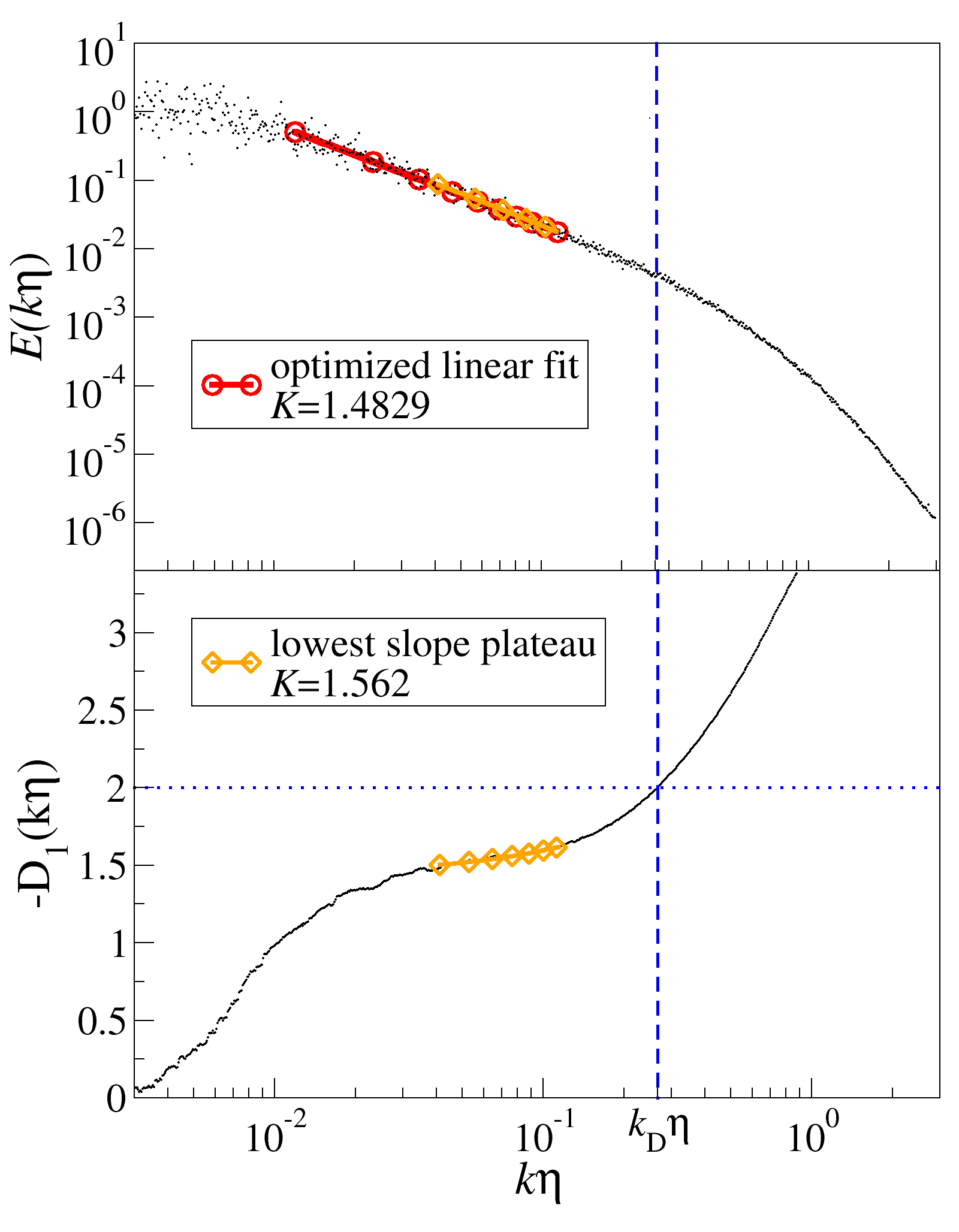}
\end{center}
\caption{\label{method1} Result of the optimized fitting algorithm for the determination of $K$ from the log-log spectrum (red line with circles)
 and from the first logarithmic derivative $D_1$ (orange line with diamonds). The result from $D_1$ is reported on the spectrum $E$
  for comparison.}
\end{figure}

 \begin{figure}[h]
\begin{center}
\includegraphics[width=6.cm]{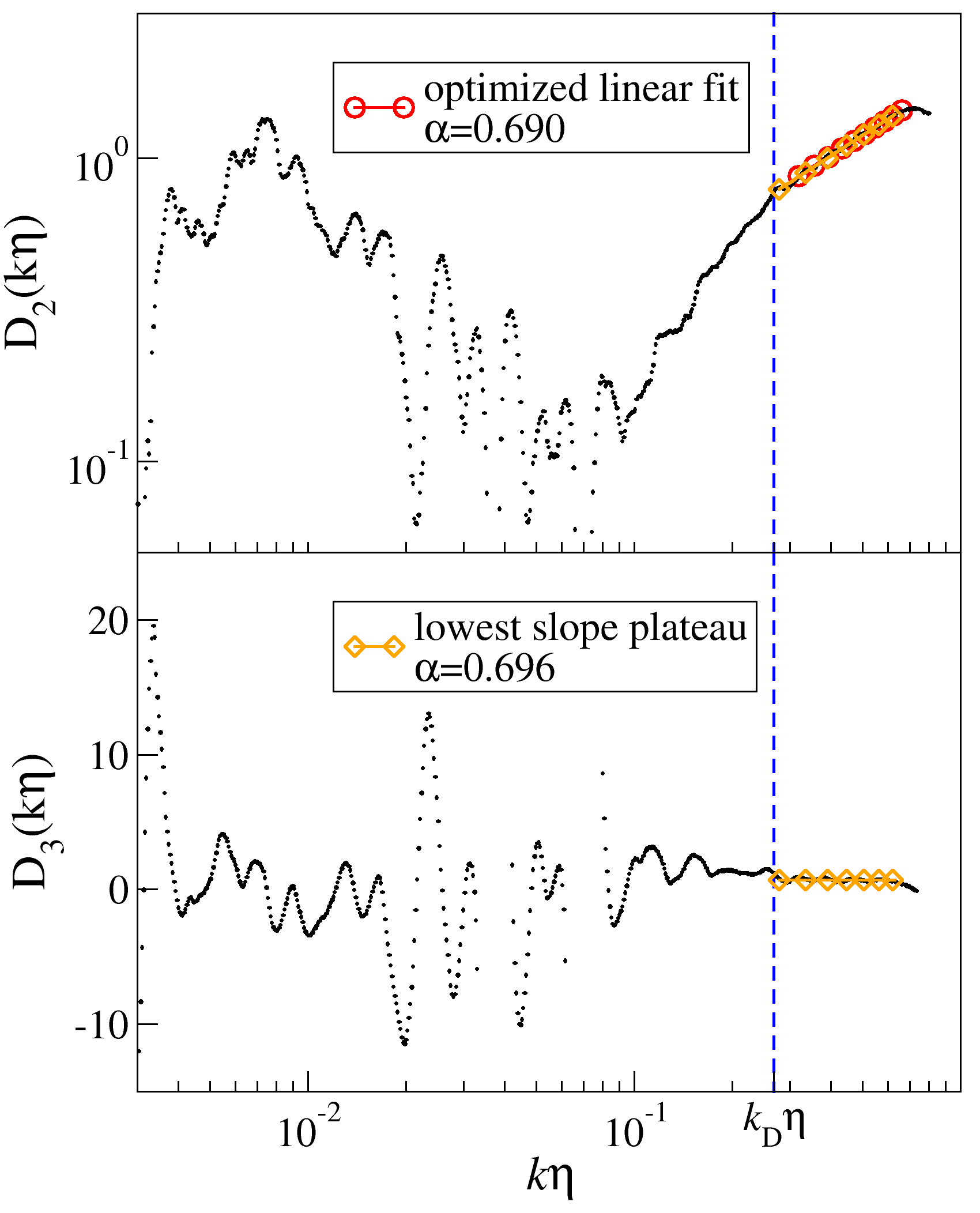}
\end{center}
\caption{\label{method2} Result of the optimized fitting algorithm for the determination of $\alpha$ from the second (red  line with circles)
 and from the third (orange line with diamonds) logarithmic derivatives of the spectrum. The result from $D_3$ is reported on the $D_2$
 curve for comparison.}
\end{figure}

As an illustration of the procedures described above, we present their result on
 a typical spectrum. The determination of $K$ is illustrated in \fref{method1},
  which shows this particular spectrum $E(k\eta)$ in log-log scale and its first logarithmic derivative
  $D_1(k\eta)$, obtained as explained in the previous section. 
  The $D_1$ curve is first used to estimate the dissipative wave-number $k_{\rm D}$,
   which we define as the wave-number for which  $(k\eta)^2 E(k\eta)$ is maximum, and hence corresponds to the value $D_1(k_{\rm D}\eta)=2$.
   We find $k_{\rm D}\eta\simeq 0.18$, which is in agreement with typical values in the literature
    and also with the value found for the numerical spectra. This value can be used as a reliable
    upper bound for the inertial range.
   The search for the largest best domain with linear behavior for $E$
    and lowest slope plateau for $D_1$ is initialized
    in the range {$k\eta\in[0.01,k_{\rm D}\eta]$}. The result of the optimized algorithm is represented, with their respective
     domains, as the red line with circles and orange line with diamonds
     respectively. In this example, the two values for $K$ differ by more than 2\% so the value from the
      linear fit of $E$ is retained since it corresponds to the largest domain.

 The determination of $\alpha$ is illustrated in \fref{method2}, which shows the second $D_2$ and third $D_3$
  logarithmic derivatives defined in \eq{der2} and \eq{der3} respectively. The numerical derivatives
   are evaluated as described in the previous section. The search domain for the near-dissipative
    range is initialized as {$k\eta\in[k_{\rm D}\eta,k_M\eta=3]$}. The result of the optimized
    linear fitting algorithm applied to  $D_2$ to extract $\alpha_2$ (and $\mu\alpha_2^2$
     as the second parameter of the fit)
    is represented by the red line with circles, the result of the lowest slope plateau algorithm applied to
      $D_3$ to extract $\alpha_3$
     is represented by the orange line with diamonds. In this example, the two results are in close agreement.


\begin{thebibliography}{10}

\bibitem{Kolmogorov41a}
A.~N. Kolmogorov, Dokl. Akad. Nauk SSSR {\bf 30},  299  (1941).

\bibitem{Kolmogorov91a}
A.~N. Kolmogorov, Proceedings of the Royal Society of London A: Mathematical,
  Physical and Engineering Sciences {\bf 434},  9  (1991).

\bibitem{Kolmogorov41c}
A.~N. Kolmogorov, Dokl. Akad. Nauk SSSR {\bf 32},  16  (1941).

\bibitem{Kolmogorov91b}
A.~N. Kolmogorov, Proceedings of the Royal Society of London A: Mathematical,
  Physical and Engineering Sciences {\bf 434},  15  (1991).

\bibitem{Kraichnan59}
R.~H. Kraichnan, Journal of Fluid Mechanics {\bf 5},  497–543  (1959).

\bibitem{Tatarskii67}
V.~I. Tatarskii, {\em {\it Line of sight propagation fluctuations}
  \textnormal{in} {\it{Atmospheric Turbulence and radio wave propagation}},
  \textnormal{edited by A. M. Yaglom and V. I. Tatarskii}}, {\em 314-329}
  (Nauka Press, Moscow, 1967).

\bibitem{Uberoi69}
M.~S. Uberoi and P. Freymuth, Physics of Fluids {\bf 12},  1359  (1969).

\bibitem{Pao65}
Y. Pao, Physics of Fluids {\bf 8},  1063  (1965).

\bibitem{Townsend51b}
A.~A. Townsend, Proceedings of the Royal Society of London A: Mathematical,
  Physical and Engineering Sciences {\bf 208},  534  (1951).

\bibitem{Novikov61}
E.~A. Novikov, Dokl. Akad. Nauk SSSR {\bf 139},  331  (1961).

\bibitem{Gurvich67}
A.~S. Gurvich, B.~M. Koprov, L.~R. Tsvang, and A.~M. Yaglom, {\em {\it Data on
  the small-scale structure of atmospheric turbulence} \textnormal{in}
  {\it{Atmospheric Turbulence and radio wave propagation}}, \textnormal{edited
  by A. M. Yaglom and V. I. Tatarskii}}, {\em 30-52} (Nauka Press, Moscow,
  1967).

\bibitem{Monin73}
A.~S. Monin and A.~M. Yaglom, {\em Statistical Fluid Mechanics: Mechanics of
  turbulence}, 2th edition ed. (MIT Press, Cambridge, Massachusetts and London,
  England, 1973).

\bibitem{Foias90}
C. Foias, O. Manley, and L. Sirovich, Physics of Fluids A: Fluid Dynamics {\bf
  2},  464  (1990).

\bibitem{Sirovich94}
L. Sirovich, L. Smith, and V. Yakhot, Phys. Rev. Lett. {\bf 74},  1492  (1995).

\bibitem{Lohse95}
D. Lohse and A. M\"uller-Groeling, Phys. Rev. Lett. {\bf 74},  1747  (1995).

\bibitem{Frisch91}
U. Frisch and M. Vergassola, Europhysics Letters ({EPL}) {\bf 14},  439
  (1991).

\bibitem{Sreenivasan85}
K.~R. Sreenivasan, Journal of Fluid Mechanics {\bf 151},  81–103  (1985).

\bibitem{Smith91}
L.~M. Smith and W.~C. Reynolds, Physics of Fluids A: Fluid Dynamics {\bf 3},
  992  (1991).

\bibitem{She93}
Z. She and E. Jackson, Physics of Fluids A: Fluid Dynamics {\bf 5},  1526
  (1993).

\bibitem{Saddoughi94}
S.~G. Saddoughi and S.~V. Veeravalli, Journal of Fluid Mechanics {\bf 268},
  333–372  (1994).

\bibitem{Sanada92}
T. Sanada and V. Shanmugasundaram, Physics of Fluids A: Fluid Dynamics {\bf 4},
   1245  (1992).

\bibitem{Chen93}
S. Chen {\it et~al.}, Phys. Rev. Lett. {\bf 70},  3051  (1993).

\bibitem{Martinez97}
D.~O. Martinez {\it et~al.}, Journal of Plasma Physics {\bf 57},  195–201
  (1997).

\bibitem{Ishihara05}
T. Ishihara {\it et~al.}, Journal of the Physical Society of Japan {\bf 74},
  1464  (2005).

\bibitem{Schumacher07}
J. Schumacher, Europhysics Letters ({EPL}) {\bf 80},  54001  (2007).

\bibitem{Ishihara09}
T. Ishihara, T. Gotoh, and Y. Kaneda, Annual Review of Fluid Mechanics {\bf
  41},  165  (2009).

\bibitem{Verma18}
M.~K. Verma {\it et~al.}, Fluid Dynamics {\bf 53},  862  (2018).

\bibitem{Manley92}
O.~P. Manley, Physics of Fluids A: Fluid Dynamics {\bf 4},  1320  (1992).

\bibitem{Pope00}
S.~B. Pope, {\em Turbulent Flows} (Cambridge University Press, Cambridge, 2000).

\bibitem{Khurshid18}
S. Khurshid, D.~A. Donzis, and K.~R. Sreenivasan, Phys. Rev. Fluids {\bf 3},
  082601  (2018).

\bibitem{Tarpin18}
M. Tarpin, L. Canet, and N. Wschebor, Physics of Fluids {\bf 30},  055102
  (2018).

\bibitem{Tarpin19}
M. Tarpin, L. Canet, C. Pagani, and N. Wschebor, Journal of Physics A:
  Mathematical and Theoretical {\bf 52},  085501  (2019).

\bibitem{Canet17}
L. Canet, V. Rossetto, N. Wschebor, and G. Balarac, Phys. Rev. E {\bf 95},
  023107  (2017).

\bibitem{Debue18}
P. Debue {\it et~al.}, Phys. Rev. Fluids {\bf 3},  024602  (2018).

\bibitem{Bourgoin17}
M. Bourgoin {\it et~al.}, 
CEAS Aeronaut. J. {\bf 9}, 269–281 (2018).

\bibitem{Dominicis79}
C. DeDominicis and P.~C. Martin, Phys. Rev. A {\bf 19},  419  (1979).

\bibitem{Fournier83}
J.~D. Fournier and U. Frisch, Phys. Rev. A {\bf 28},  1000  (1983).

\bibitem{Smith98}
L. Smith and S. Woodruff, Annu. Rev. Fluid Mech. {\bf 30},  275  (1998).

\bibitem{Adzhemyan99}
L.~T. Adzhemyan, N.~V. Antonov, and A.~N. Vasil'ev, {\em The Field Theoretic
  {R}enormalization {G}roup in Fully Developed Turbulence} (Gordon and Breach,
  London, 1999).

\bibitem{Zhou10}
Y. Zhou, Phys. Rep. {\bf 488},  1   (2010).

\bibitem{Wilson74}
K.~G. Wilson and J. Kogut, Phys. Rep. C {\bf 12},  75  (1974).

\bibitem{Berges02}
J. Berges, N. Tetradis, and C. Wetterich, Phys. Rep. {\bf 363},  223   (2002).

\bibitem{Kopietz10}
P. Kopietz, L. Bartosch, and F. Sch\"utz, {\em {Introduction to the Functional
  Renormalization Group}}, {\em Lecture Notes in Physics} (Springer, Berlin,
  2010).

\bibitem{Delamotte12}
B. Delamotte, {\em {\it An introduction to the Nonperturbative Renormalization
  Group} \textnormal{in} {\it{Renormalization Group and Effective Field Theory
  Approaches to Many-Body Systems}}, \textnormal{edited by J. Polonyi and A.
  Schwenk}}, {\em Lecture Notes in Physics} (Springer, Berlin, 2012).

\bibitem{Tomassini97}
P. Tomassini, Phys. Lett. B {\bf 411},  117  (1997).

\bibitem{Monasterio12}
C. Mej\'ia-Monasterio and P. Muratore-Ginanneschi, Phys. Rev. E {\bf 86},
  016315  (2012).

\bibitem{Canet16}
L. Canet, B. Delamotte, and N. Wschebor, Phys. Rev. E {\bf 93},  063101
  (2016).

\bibitem{Note1}
\modif{Let us emphasize that the precise profile chosen for $N$ is not important as it
  does not influence the universal properties of the flow, as was shown in
  \cite {Tomassini97}. It can also be chosen diagonal in component space,
  without loss of generality because of incompressibility \cite {Canet16}.
  Moreover, although one may argue that a forcing uncorrelated in time is not
  realistic physically, it was shown that it plays no role for the universal
  properties. Indeed, introducing finite time correlations in (\ref
  {variance_f}) does not alter the universal properties, as long as these
  correlations are not too long-ranged, as was shown in \cite {Antonov18} for
  Navier-Stokes equation with a power-law forcing and in \cite {Squizzato19}
  for Burgers equation with both short-range and power-law forcing.}

\bibitem{Martin73}
P.~C. Martin, E.~D. Siggia, and H.~A. Rose, Phys. Rev. A {\bf 8},  423  (1973).

\bibitem{Janssen76}
H.-K. Janssen, Z. Phys. B {\bf 23},  377  (1976).

\bibitem{Dominicis76}
C. de~Dominicis, J. Phys. (Paris) Colloq. {\bf 37},  247  (1976).

\bibitem{Canet15}
L. Canet, B. Delamotte, and N. Wschebor, Phys. Rev. E {\bf 91},  053004
  (2015).

\bibitem{Tennekes75}
H. Tennekes, J. Fluid Mech. {\bf 67},  561  (1975).

\bibitem{Lagaert14}
J.-B. Lagaert, G. Balarac, and G.-H. Cottet, J. Comp. Phys. {\bf 260},
  (2014).

\bibitem{Falkovich94c}
G. Falkovich, Physics of Fluids {\bf 6},  1411  (1994).

\bibitem{Donzis10}
D.~A. Donzis and K.~R. Sreenivasan, Journal of Fluid Mechanics {\bf 657},
  171–188  (2010).

\bibitem{Antonov18}
N.~V. Antonov, N.~M. Gulitskiy, M.~M. Kostenko, and A.~V. Malyshev, Phys. Rev.
  E {\bf 97},  033101  (2018).

\bibitem{Squizzato19}
D. Squizzato and L. Canet, Phys. Rev. E {\bf 100},  062143  (2019).

\end{thebibliography}

\end{document}